\documentclass[onecolumn,aps,prd,preprintnumbers,showpacs,superscriptaddress,nofootinbib,amsmath,amssymb,floats,floatfix,showkeys,notitlepage,longbibliography]{revtex4-1}

\usepackage{graphicx}
\usepackage{subfigure}
\usepackage{palatino}
\usepackage[toc,page]{appendix}
\usepackage[normalem]{ulem}
\usepackage{sans}
\usepackage{hyperref}
\hypersetup{colorlinks=true,linkcolor=blue,urlcolor=blue,citecolor=blue}
\usepackage[normalem]{ulem}
\usepackage{adjustbox}
\usepackage{latexsym}
\usepackage{amsmath}
\usepackage{amssymb}
\usepackage{amsfonts}
\usepackage{dcolumn}
\usepackage{bm}
\usepackage{bigints}
\usepackage{array,tabularx,multirow,booktabs}
\UseRawInputEncoding 
\allowdisplaybreaks
\usepackage{color}

\def\0{{\sst{(0)}}}
\def\1{{\sst{(1)}}}
\def\2{{\sst{(2)}}}
\def\3{{\sst{(3)}}}
\def\4{{\sst{(4)}}}
\def\5{{\sst{(5)}}}
\def\6{{\sst{(6)}}}
\def\7{{\sst{(7)}}}
\def\8{{\sst{(8)}}}
\def\sst#1{{\scriptscriptstyle #1}}

\begin{document} \sloppy

\title{Testing Symmergent gravity through the shadow image and weak field photon deflection by a rotating black hole using the M87$^*$ and Sgr. A$^*$ results}

\author{Reggie C. Pantig}
\email{rcpantig@mapua.edu.ph}
\affiliation{Physics Department, Map\'ua University, 658 Muralla St., Intramuros, Manila 1002, Philippines}

\author{Ali \"Ovg\"un}
\email{ali.ovgun@emu.edu.tr}
\homepage[]{https://aovgun.com}
\affiliation{Physics Department, Eastern Mediterranean
University, Famagusta, 99628 North Cyprus, via Mersin 10, Turkey}

\author{Durmu\c{s}~Demir}
\email{durmus.demir@sabanciuniv.edu}
\affiliation{Faculty of Engineering and Natural Sciences, Sabanc{\i} University, 34956 Tuzla, \.{I}stanbul, Turkey}

\begin{abstract}
In this paper, we study rotating black holes in symmergent gravity, and use deviations from the Kerr black hole to constrain the parameters of the symmergent gravity. Symmergent gravity induces the gravitational constant $G$ and quadratic curvature coefficient $c_{\rm O}$ from the flat spacetime matter loops. In the limit in which all fields are degenerate in mass, the vacuum energy $V_{\rm O}$ can be wholly expressed in terms of $G$ and $c_{\rm O}$. We parametrize deviation from this degenerate  limit by a parameter ${\hat \alpha}$ such that the black hole spacetime is dS for ${\hat \alpha} < 1$ and AdS for  ${\hat \alpha} > 1$. In constraining the symmergent parameters $c_{\rm O}$ and ${\hat \alpha}$, we utilize the EHT observations on the M87* and Sgr. A* black holes. We  investigate first the modifications in the photon sphere and shadow size, and find significant deviations in the photonsphere radius and the shadow radius with respect to the Kerr solution. We also find that the geodesics of time-like particles are more sensitive to symmergent gravity effects than the null geodesics. Finally, we analyze the weak field limit of the deflection angle, where we use the Gauss-Bonnet theorem for taking into account the finite distance of the source and the receiver to the lensing object. Remarkably, the distance of the receiver (or source) from the lensing object greatly influences the deflection angle. Moreover,   $c_{\rm O}$ needs  be negative for a consistent solution. In our analysis, the rotating black hole acts as a particle accelerator and possesses the sensitivity to probe the symmergent gravity.
\end{abstract}

\date{\today}



\maketitle

\section{Introduction}

In the Wilsonian sense, quantum field theories (QFTs) are characterized by a classical action and an ultraviolet (UV) cutoff $\Lambda$. Quantum loops lead to effective QFTs with loop momenta cut at $\Lambda$. The effective QFTs suffer from UV oversensitivity problems: The scalar and gauge boson masses receive ${\mathcal{O}}(\Lambda^2)$ corrections. The vacuum energy, on the other hand, gets corrected by ${\mathcal{O}}(\Lambda^4)$  and ${\mathcal{O}}(\Lambda^2)$  terms. The gauge symmetries get explicitly broken. The question is simple: Can gravity emerge in a way restoring the explicitly broken gauge symmetries? Asking differently, can gravity emerge in a way alleviating the UV oversensitivities of the effective QFT? This question has been answered affirmatively  by forming a gauge symmetry-restoring emergent gravity model \cite{demir1,demir2,demir3}. This model, briefly called as {\it symmergent gravity}, has been built by the observation that, in parallel with the introduction of Higgs field to restore gauge symmetry for a massive vector boson (with Casimir invariant mass) \cite{anderson,englert,higgs}, spacetime affine curvature can be  introduced to restore gauge symmetries for gauge bosons with loop-induced (Casimir non-invariant) masses proportional to the UV cutoff $\Lambda$ \cite{demir1,demir2,demir3}. Symmergent gravity is essentially emergent general relativity (GR) with a quadratic curvature term. It exhibits distinctive signatures, as revealed in recent works on static black hole spacetimes \cite{irfan,Symmergent-bh,Symmergent-bh2,symmergentresults}. In the present work, we use observational data on M87* and Srg.A* black holes to study symmergent gravity. We focus on rotating black hole spacetimes as these black holes can act as a richer test and analysis laboratory. What we are doing can  be viewed as ``doing particle physics via black holes" as it will reveal salient properties of the QFT sector through various black hole features.

In 1919 Arthur Eddington led an expedition to prove Einstein's theory of relativity using gravitational lensing and since then lensing has become an important tool in astrophysics \cite{Virbhadra:1999nm,Virbhadra:2002ju,Virbhadra:1998dy,Virbhadra:2007kw,Virbhadra:2008ws,Adler:2022qtb,Bozza:2001xd,Bozza:2002zj,Perlick:2003vg,He:2020eah}.  In astrophysics, distances are very important when determining the properties of astrophysical objects. But Virbhadra showed that just the observation of relativistic images without any information about the masses and distances can also accurately give value of the upper bound on the compactness of massive dark objects \cite{Virbhadra:2022ybp}. Furthermore, Virbhadra proved that there exists a distortion parameter such that the signed sum of all images of singular gravitational lensing of a source identically vanishes (tested with Schwarzschild lensing in weak and strong gravitational fields \cite{Virbhadra:2022iiy}).

In 2008, Gibbons and Werner used the Gauss-Bonnet theorem (GBT) on the optical geometries in asymptotically flat spacetimes, and calculated weak deflection angle for first time in literature \cite{Gibbons:2008rj}. Afterwards, this method has been applied to various  phenomena  \cite{Ovgun:2018fnk,Ovgun:2019wej,Ovgun:2018oxk,Javed:2019ynm,Werner_2012,Ishihara:2016vdc,Ishihara:2016sfv,Ono:2017pie,Li:2020dln,Li:2020wvn,Belhaj:2022vte, Javed:2023iih, Javed:2023IJGMMP,Pantig:2022ely}. One of the aims of the present paper is to probe symmergent gravity  through the black hole's weak deflection angle using the GBT and also the shadow silhouette as perceived by a static remote observer. In essence, by the very nature of the symmergent gravity, we will be studying the numbers of fermions and bosons and related effects in the vicinity of a spinning black hole, which has the potential to affect the motions of null and time-like geodesics as well as their spin parameter $a$. (By definition, $a=c J/G M^2$ where $J$ and $M$ are the angular momentum and mass of the rotation black hole.) Recently,  shadow coming from different black hole models has been analyzed extensively in the literature as a probe of the imprints of the astrophysical environment affecting it \cite{Vagnozzi:2022moj,Chen:2022nbb,Dymnikova2019,Pantig:2022whj,Uniyal:2022vdu,Kuang:2022xjp,Meng:2022kjs,Tang:2022hsu,Kuang:2022ojj,Wei2019,Xu2018a,Hou:2018avu,Bambi2019,Tsukamoto:2017fxq,Kumar:2020hgm,Kumar2019,Wang2017,Wang2018,Amarilla2018,He2020,Tsupko_2020,Hioki2009,Konoplya2019,Okyay:2021nnh,Belhaj:2020rdb,Li2020,Ling:2021vgk,Belhaj:2020okh,Cunha:2018acu,Gralla:2019xty,Perlick:2015vta,Nedkova:2013msa,Li:2013jra,Khodadi:2021gbc,Khodadi:2022pqh,Cunha:2016wzk,Shaikh:2019fpu,Allahyari:2019jqz,Yumoto:2012kz,Cunha:2016bpi,Moffat:2015kva,Cunha:2016bjh,Zakharov:2014lqa,Hennigar:2018hza,Chakhchi:2022fl, Pantig:2022sjb, Pantig:2022gih,Lobos:2022jsz,Pantig:2022toh}
Calculation of the shadow cast of a non-rotating black hole was pioneered by \cite{Synge1966} and \cite{Luminet1979}, and later on extended by \cite{Bardeen:1973tla} \cite{Chandrasekhar98} to an axisymmetric spacetime. Recently, after the detection of gravitational waves in 2015 \cite{LIGOScientific:2016aoc}, the first image of the black hole in M87* was formed by the Event Horizon Telescope (EHT) using its electromagnetic spectrum \cite{EventHorizonTelescope:2019ggy}. At the time of writing of this paper, new milestone has been achieved since the EHT revealed the shadow image of the black hole in our galaxy, Sgr. A* \cite{EventHorizonTelescope:2022xnr}. This new result indicates that black hole theory and phenomenology is a hot research topic in vivew of the rapidly developing observational techniques.

The paper is organized as follows. In Sec. \ref{sec2}, we give a detailed discussion of the symmergent gravity. We give in Sec. \ref{sec3} the rotating metric, indicating dependence on the symmergent parameters. In Sec. \ref{sec4}, we study the effects of symmergent gravity on null geodesics, which in turn influence the shadow radius and observables associated with it. We also examine the said effect on time-like orbits in the same section. We devote  Sec. \ref{sec5}  to investigation of the weak deflection angle at finite-distance, and in Sec. \ref{sec6} we calculate the center-of-mass energy (CM) of two particles and study particle acceleration near rotating symmergent black hole background. In Sec. \ref{sec7}, we conclude and give future prospects.  

\section{Symmergent Gravity} \label{sec2}

In this section, we give a brief description of the symmergent gravity in terms of its fundamental parameters. The starting point is quantum field theories (QFTs). Quantum fields are endowed with mass and spin as the Casimir invariant s of the Poincar\'e group. Fundamentally, QFTs are intrinsic to the flat spacetime simply because they rest on a Poincar\'e-invariant (translation-invariant) vacuum state \cite{incompatible,wald}.  Flat spacetime means the total absence of gravity. Incorporation of gravity necessitates the QFTs to be taken to curved spacetime, but this is hampered by Poincar\'e breaking in curved spacetime \cite{dyson,wald}. This hamper and absence of a quantum theory of gravity \cite{thooft}  together lead one to emergent gravity framework \cite{sakharov,visser,verlinde} as a viable approach.

In general, loss of Poincar\'e invariance could be interpreted as the emergence of gravity into the QFT \cite{fn2}. In a QFT, curvature can emerge at the Poincar\'e breaking sources. One natural Poincar\'e breakings source  is the hard momentum cutoff on the QFT. Indeed, an ultraviolet (UV)  cutoff $\Lambda$ \cite{cutoff} limits momenta $p_\mu$ within $-\Lambda^2 \leq \eta^{\mu\nu} p_\mu p_\nu \leq \Lambda^2$ interval as the intrinsic validity edge of the QFT \cite{cutoff}. Under the loop corrections up to the cutoff $\Lambda$, the  action $S[\eta,\phi,V]$ of a QFT of scalars $S$ and gauge bosons $V_\mu$ receives the correction (with $(+,-,-,-)$ metric signature appropriate for QFTs)
\begin{eqnarray}
 \delta S[\eta,\phi,V]  = \int d^4 x \sqrt{-\eta} \left\{-V_{\rm O} -c_{\rm O} \Lambda^4 - \sum_i c_{m_i} m_i^2 \Lambda^2 - c_S \Lambda^2 S^\dagger S + c_V \Lambda^2 V_{\mu} V^{\mu} \right\}
 \label{action-0}
\end{eqnarray}
in which $\eta_{\mu\nu}$ is the flat metric, $V_{\rm O}$ is the vacuum energy which is not power-law in $\Lambda$,  $m_i$ stands for the mass of a QFT field $\psi_i$ (summing over all the fermions and bosons),  and $c_{\rm O}$, $c_m$, $c_S$ and $c_V$ are respectively the loop factors describing the quartic vacuum energy correction, quadratic vacuum energy correction, quadratic scalar mass correction, and the loop-induced gauge boson mass \cite{gauge-break1,gauge-break2}. As revealed by the gauge boson mass term $c_V \Lambda^2 V_{\mu} V^{\mu}$, the UV cutoff $\Lambda$ breaks gauge symmetries explicitly since $\Lambda$ is not a particle mass, that is, $\Lambda$ is not a Casimir invariant of the Poincar\'e group. The loop factor $c_V$ (as well as $c_S$) depends on the details of the QFT. (It has been calculated for the standard model gauge group in \cite{demir1,demir2}.)

In Sakharov's induced gravity \cite{sakharov,visser}, the UV cutoff $\Lambda$ is associated with the Planck scale, albeit with explicitly-broken gauge symmetries and Planckian-size cosmological constant and scalar masses. In recent years, Sakharov's setup has been approached from a new perspective in which priority is given to the prevention of the explicit gauge symmetry breaking \cite{demir1}. For this aim, one first takes the effective QFT in (\ref{action-0}) to curved spacetime of a metric $g_{\mu\nu}$ such that the gauge boson mass term is mapped as $c_V \Lambda^2 \eta^{\mu\nu} V_{\mu} V_{\nu} \longrightarrow c_V V_{\mu} \left(\Lambda^2 g^{\mu\nu}-R^{\mu\nu}(g)\right)V_\nu$ in agreement with the fact that the Ricci curvature $R^{\mu\nu}(g)$ of the metric $g_{\mu\nu}$ can arise only in the gauge sector via the covariant derivatives \cite{demir1,demir2,demir3}. One next inspires from the Higgs mechanism to promote the UV cutoff $\Lambda$ to an appropriate spurion field. Indeed, in parallel with the introduction of Higgs field to restore gauge symmetry for a massive vector boson (Poincare-conserving mass) \cite{anderson,englert,higgs}, one can introduce spacetime affine curvature  to restore gauge symmetries for gauge bosons with loop-induced (Poincare-breaking mass) masses proportional to $\Lambda$ \cite{affine1,affine2,demir1,demir2}. Then, one is led to the map \begin{eqnarray}
\Lambda^2 g^{\mu\nu} \rightarrow {\mathbb{R}}^{\mu\nu}(\Gamma)
\label{map}
\end{eqnarray}
in which ${\mathbb{R}}^{\mu\nu}(\Gamma)$ is the Ricci curvature of the affine connection $\Gamma^{\lambda}_{\mu\nu}$, which is completely independent of the curved metric $g_{\mu\nu}$ and its Levi-Civita connection \cite{affine1,affine2,affine3}. This map is analogue of the map $M_V^2 \rightarrow \phi^\dagger \phi$ of the vector boson mass $M_V$ (Poincare conserving) into the Higgs field $\phi$. Under the map (\ref{map}) the effective QFT in (\ref{action-0}) takes the form
\begin{eqnarray}
 \delta S[g,\phi,V, {\mathbb{R}}]  = \int d^4 x \sqrt{-\eta} \left\{-V_{\rm O}-\frac{c_{\rm O}}{16} {\mathbb{R}}^2(g) - \sum_i \frac{c_{m_i}}{4} m_i^2 {\mathbb{R}}(g) - \frac{c_S}{4} {\mathbb{R}}(g) S^\dagger S + c_V V_{\mu} \left({\mathbb{R}}^{\mu\nu}(\Gamma)-R^{\mu\nu}(g)\right)V_\nu \right\}
 \label{action-1}
\end{eqnarray}
in which ${\mathbb{R}}(g)\equiv g^{\mu\nu} {\mathbb{R}}_{\mu\nu}(\Gamma)$ is the scalar affine curvature \cite{demir1}. This metric-Palatini theory  contains  both the metrical curvature $R(g)$ and the affine curvature ${\mathbb{R}}(\Gamma)$. From the third term, Newton's gravitational constant $G$ is read out to be 
\begin{eqnarray}
G^{-1}= 4 \pi \sum_i c_{m_i} m_i^2 \xrightarrow{\rm one\ loop} \frac{1}{8\pi} {\rm str}\!\left[{\mathcal{M}}^2 \right]
\label{MPl}
\end{eqnarray}
where ${\mathcal{M}}^2$ is the mass-squared matrix of all the fields in the QFT spectrum. In the one-loop expression, ${\rm str}[\dots]$ stands for super-trace namely ${\rm str}[{\mathcal{M}}^2] = \sum_i (-1)^{2s_i} (2 s_i +1) {\rm tr}[{\mathcal{M}}^2]_{s_i}$ in which ${\rm tr}[\dots]$
is the usual trace (including the color degrees of freedom), $s_i$ is the spin of the QFT field $\psi_i$  ($s_i=0,1/2,\dots$), and $[{\mathcal{M}}^2]_{s_i}$ is the mass-squared matrix of the fields having that spin (like mass-squared matrices of scalars ($s_i=0$), fermions $(s_i=1/2)$ and so on). One keeps in mind that ${\rm tr}[\dots]$ encodes degrees of freedom $g_i$ (like color and other degrees of freedom) of the particles. 

It is clear that known particles (the standard model spectrum) cannot generate Newton's constant in (\ref{MPl}) correctly (in both sign and size). It is necessary to introduce therefore new particles. Interesting enough, these new particles do not have to couple to the known particles since the only constraint on them is the super-trace in (\ref{MPl}) \cite{demir1,demir2}. 

The action (\ref{action-1}) remains stationary against variations in the affine connection provided that 
\begin{eqnarray}
\label{gamma-eom}
{}^{\Gamma}\nabla_{\lambda} {\mathbb{D}}_{\mu\nu} = 0
\end{eqnarray}
such that ${}^{\Gamma}\nabla_{\lambda}$ is the covariant derivative of the affine connection $\Gamma^\lambda_{\mu\nu}$, and 
\begin{eqnarray}
\label{q-tensor}
{\mathbb{D}}_{\mu\nu} = \left(\frac{1}{16\pi G} +   \frac{c_S}{4} S^\dagger S + \frac{c_{\rm O}}{8} g^{\alpha\beta} {\mathbb{R}}_{\alpha\beta}(\Gamma)\right) g_{\mu\nu} - c_{V} V_{\mu}V_{\nu}
\end{eqnarray}
is the field-dependent metric.  The motion equation (\ref{gamma-eom}) implies that ${\mathbb{D}}_{\mu\nu}$ is covariantly-constant with respect to $\Gamma^\lambda_{\mu\nu}$, and this constancy leads to the exact  solution
\begin{eqnarray}
\Gamma^\lambda_{\mu\nu} = \frac{1}{2} ({\mathbb{D}}^{-1})^{\lambda\rho} \left( \partial_\mu {\mathbb{D}}_{\nu\rho} + \partial_\nu {\mathbb{D}}_{\rho\mu} - \partial_\rho {\mathbb{D}}_{\mu\nu}\right)={}^g\Gamma^\lambda_{\mu\nu} + \frac{1}{2} ({\mathbb{D}}^{-1})^{\lambda\rho} \left( \nabla_\mu {\mathbb{D}}_{\nu\rho} + \nabla_\nu {\mathbb{D}}_{\rho\mu} - \nabla_\rho {\mathbb{D}}_{\mu\nu}\right)
\label{aC}
\end{eqnarray}
in which ${}^g\Gamma^\lambda_{\mu\nu}$ is the Levi-Civita connection of the curved metric $g_{\mu\nu}$. The Planck scale in (\ref{MPl}) is the largest scale and therefore it is legitimate to make the expansions 
\begin{eqnarray}
\Gamma^{\lambda}_{\mu\nu}&=&{}^{g}\Gamma^{\lambda}_{\mu\nu} + 8\pi G \left( \nabla_\mu {\mathbb{D}}^\lambda_\nu + \nabla_\nu {\mathbb{D}}^\lambda_\mu - \nabla^\lambda {\mathbb{D}}_{\mu\nu}\right) + {\mathcal{O}}\left(G^2\right)
\label{expand-conn}
\end{eqnarray}
and
\begin{eqnarray}
{\mathbb{R}}_{\mu\nu}(\Gamma) &=& R_{\mu\nu}(g) + 8\pi G\left(\nabla^{\alpha} \nabla_{\mu} {\mathbb{D}}_{\alpha\nu} + \nabla^{\alpha} \nabla_{\nu} {\mathbb{D}}_{\alpha\mu} - \Box {\mathbb{D}}_{\mu\nu} - \nabla_{\mu} \nabla_{\nu} {\mathbb{D}}_{\alpha}^{\alpha}\right)  +  {\mathcal{O}}\left(G^2\right)
\label{expand-curv}
\end{eqnarray}
so that both $\Gamma^{\lambda}_{\mu\nu}$ and ${\mathbb{R}}_{\mu\nu}(\Gamma)$ contain  pure derivative terms  at  
the next-to-leading ${\mathcal{O}}\left(G\right)$ order \cite{demir2,demir3}. The expansion in (\ref{expand-conn}) ensures that the affine connection $\Gamma^{\lambda}_{\mu\nu}$ is solved algebraically order by order in $G$  despite the fact that its motion equation (\ref{gamma-eom}) involves its own curvature ${\mathbb{R}}_{\mu\nu}(\Gamma)$ through ${\mathbb{D}}_{\mu\nu}$   \cite{affine1,affine2}. The expansion (\ref{expand-curv}), on the other hand, ensures that the affine curvature  ${\mathbb{R}}_{\mu\nu}(\Gamma)$ is equal to the metrical curvature $R_{\mu\nu}(g)$ up to a doubly-Planck suppressed remainder. In essence, what happened is that the affine dynamics took the affine curvature ${\mathbb{R}}$ from its UV value $\Lambda_\wp^2$  in (\ref{map}) to its IR value $R$ in (\ref{expand-curv}). This way, the GR emerges holographically \cite{holog1,holog2} via the affine dynamics such that loop-induced gauge boson masses get erased, and scalar masses get stabilized by the curvature terms. This mechanism renders effective field theories natural regarding their destabilizing UV sensitivities \cite{demir1,demir2}. It gives rise to a new framework in which {\it (i)} the gravity sector is composed of the Einstein-Hilbert term plus a curvature-squared term, and {\it (ii)} the matter sector is described by an ${\overline{MS}}$-renormalized QFT \cite{demir1,demir2}. We call this framework gauge symmetry-restoring emergent gravity or simply {\it symmergent gravity} to distinguish is from other emergent or induced gravity theories in the literature. 

It is worth noting that symmergent gravity is not a loop-induced curvature sector in curved spacetime  \cite{visser,birrel}. In contrast, symmergent gravity arises when the flat spacetime effective QFT is taken to curved spacetime \cite{demir1,demir2} in a way reviving the gauge symmetries broken explicitly by the UV cutoff. All of its couplings are loop-induced parameters deriving from the particle spectrum of the QFT (numbers and masses of particles). It is with these loop features that the GR 
emerges. In fact, the metric-Palatini action (\ref{action-1}) reduces to the metrical gravity theory 
\begin{eqnarray}
     \int d^4x \sqrt{-g} \Bigg\{\! && - V_{\rm O}
 -\frac{{\mathbb{R}}(g)}{16\pi G} - \frac{c_{\rm O}}{16} \left({\mathbb{R}}(g)\right)^2 +\frac{c_S}{4}  S^\dagger S\, {\mathbb{R}}(g) + c_V {\rm tr}\left[V^{\mu}\!\left({\mathbb{R}}_{\mu\nu}(\Gamma)- R_{\mu\nu}({}^g\Gamma)\right)\!V^{\nu}\right]\! \Bigg\}\nonumber\\
&&\xrightarrow{\rm equation\, (\ref{expand-curv})}
\int d^4x \sqrt{-g} \Bigg\{\!
- V_{\rm O}-\frac{R}{16\pi G}   - \frac{c_{\rm O}}{16} R^2  -\frac{c_S}{4} S^\dagger S  R + {\mathcal{O}}\!\left(G\right)\!\Bigg\}
\label{reduce-nongauge}
\end{eqnarray}
after replacing the affine curvature ${\mathbb{R}}_{\mu\nu}(\Gamma)$ with its solution in (\ref{expand-curv}). Of the parameters of this emergent GR action, Newton's constant $G$ was already defined in (\ref{MPl}). The loop factor $c_S$ depends on the underlying QFT. (It reads $c_S\simeq 0.29$ in the standard model.) The loop factor $c_{\rm O}$, which was associated with the quartic ($\Lambda^4$) corrections in the flat spacetime effective QFT in (\ref{action-0}), turned to the coefficient of quadratic-curvature $(R^2)$ term in the symmergent GR action in (\ref{reduce-nongauge}). At one loop, it takes the value
\begin{eqnarray}
\label{param1}
 c_{\rm O} = \frac{n_\text{B} - n_\text{F}}{128 \pi^2}
\end{eqnarray}
in which  $n_\text{B}$ ($n_\text{F}$) stands for the total number of bosonic (fermionic) degrees of freedom in the underlying QFT (including the color degrees of freedom). Both the $n_\text{B}$ bosons and $n_\text{F}$ fermions contain not only the known standard model particles  but also the completely new particles. As was commented just above (\ref{gamma-eom}), it is a virtue of symmergence that these new particles do not have to couple to the known ones, non-gravitationally. 

The last parameter of the symmergent GR action (\ref{reduce-nongauge}) is the vacuum energy density $V_{\rm O}$. It belongs to the non-power-law sector of the flat spacetime effective QFT in (\ref{action-0}). At one loop,  it takes value (${\rm tr}[\dots]$ involves all degrees of freedom $g_i$ of particles, like color)
\begin{eqnarray}
\label{param2}
 V_{\rm O} = \frac{{\rm str}\left[{\mathcal{M}}^4\right]}{64 \pi^2}
\end{eqnarray}
after discarding a possible tree-level contribution. Being a loop-induced quantity,  Newton's constant in (\ref{MPl}) involves super-trace of  $({\rm masses})^2$ of the QFT fields. In this regard, the potential energy $V_{\rm O}$, involving the super-trace of  $({\rm masses})^4$ of the QFT fields, is expected to expressible in terms of $G$. To see this, it proves useful to start with mass degeneracy limit in which each and every boson and fermion possess equal masses, $m_b=m_f=M_0$, for all $b$ and $f$. Needless to say, $M_0$ is essentially the characteristic scale of the QFT. (Essentially, $M_0$ is the mean value of all the field masses.) Under this degenerate mass spectrum the potential $V_{\rm O}$ can be expressed as follows:
\begin{eqnarray}
\label{analyze-VO-1}
V_{\rm O} = \frac{{\rm str}\left[{\mathcal{M}}^4\right]}{64 \pi^2} = \frac{1}{64\pi^2}\left(\sum_{\rm B} m^4_{\rm B}- \sum_{\rm F} m^4_{\rm F}\right)\xrightarrow{\rm mass\,  degeneracy}\frac{M_0^4}{64\pi^2}(n_\text{B} - n_\text{F})=\frac{M_0^2}{8\pi G}=\frac{1}{2(8\pi G)^2 c_{\rm O}}
\end{eqnarray}
where use have been made of the $G$ formula in (\ref{MPl}) and $c_{\rm O}$ formula in (\ref{param1}). Now, the problem is to take into account realistic cases in which the QFT fields are not all mass-degenerate. For a QFT with characteristic scale $M_0$ but with no detailed knowledge of the mass spectrum, realistic cases might be represented by parametrizing the potential energy as
\begin{eqnarray}
\label{analyze-VO-2}
V_{\rm O} &=& \frac{1-{\hat \alpha}}{(8\pi G)^2 c_{\rm O}} 
\end{eqnarray}
in which the new parameter ${\hat \alpha}$ is introduced as a measure of the deviations of the boson and fermion masses from the QFT characteristic scale $M_0$. Clearly, ${\hat \alpha}=1/2$ corresponds to the degenerate case in (\ref{analyze-VO-1}). Alternatively,  ${\hat \alpha}=1$ represents the case in which $\sum_{\rm B} m^4_{\rm B}= \sum_{\rm F} m^4_{\rm F}$ in (\ref{analyze-VO-1}). In general,  ${\hat \alpha}>1$ (${\hat \alpha}<1$) corresponds to the boson (fermion) dominance in terms of the trace $({\rm masses})^4$. 

Symmergence makes gravity emerge from within the flat spacetime effective QFT. Fundamentally, as follows from the action (\ref{action-0}), Newton's constant $G$ in (\ref{MPl}), the quadratic curvature coefficient $c_{\rm O}$ in (\ref{param1}), and the vacuum energy $V_{\rm O}$ in (\ref{param2}) transpired in the flat spacetime effective QFT from the matter loops. (The loop factor $c_S$ and similar parameters couple the matter fields.) Once the gravity emerges as in (\ref{reduce-nongauge}), however, $G$, $c_{\rm O}$  and $V_{\rm O}$ gain a completely new physical meaning as the curvature sector parameters (not involving the matter fields). In this sense, they are symmergent gravity parameters in the sense that they became curvature sector parameters via the symmergence.

A  glance at the second line of (\ref{reduce-nongauge}) reveals that symmergent gravity is an $R+R^2$ gravity theory with non-zero cosmological constant. In fact, it can be put in the form  
\begin{eqnarray}
S=\frac{1}{16 \pi G} \int d^4 x \sqrt{-g}\left(R+f(R)\right)
\label{fr-action}
\end{eqnarray}
after leaving aside the scalars $S$ and the other matter fields, after switching to $(-,+,+,+)$ metric signature (appropriate for the black hole analysis in the sequel), and after introducing the $f(R)$ gravity function 
\begin{eqnarray}
f(R) =- \pi G c_{\rm O} R^2 -16 \pi G V_{\rm O}
\label{f(R)}
\end{eqnarray}
comprising the vacuum energy $V_{\rm O}$ and the quadratic-curvature term proportional to $c_{\rm O}$. The Einstein field equations arising from 
the action (\ref{fr-action}) 
\begin{eqnarray} 
\label{f1}
\mathit{R}_{\mu \nu} (1+f^\prime(R))-\frac{1}{2} g_{\mu \nu} (R + f(R)) + (\nabla_\mu \nabla_\nu - \Box g_{\mu\nu})f^\prime(R) = 0
\end{eqnarray}
become $R_0(f^\prime(R_0)-1) - 2f(R_0) = 0$  upon contraction at constant curvature ($R=R_0$), and possess the solution \cite{Cembranos:2011sr,Dastan:2016vhb}
\begin{eqnarray}
R_0 = 32\pi G V_{\rm O} = \frac{1-{\hat \alpha}}{2\pi G c_{\rm O}}
\label{R0-gecici}
\end{eqnarray}
after using the vacuum energy formula in (\ref{analyze-VO-2}) in the second equality. It is clear that this $R_0$ value  would vanish if the vacuum energy were not nonzero. Indeed, for a quadratic gravity with $f(R)= b R^2$ one gets the solution $R_0=0$. But for a quadratic gravity like $f(R)=\hat a+bR^2$ one finds $R_0=-2 \hat a \neq 0$, which reduces to the $R_0$ in (\ref{R0-gecici}) for $\hat a=-16 \pi G V_{\rm O}$. 

The Einstein field equations (\ref{f1}) possess 
static, spherically-symmetric, constant-curvature solutions of the form  
\cite{Cembranos:2011sr,Dastan:2016vhb}
\begin{eqnarray}
ds^2=-h(r) dt^2 + \frac{dr^2}{h(r)} + r^2 (d\theta^2+ \sin^2\theta d\phi^2)
\end{eqnarray}
in which
\begin{eqnarray} 
\label{smetric}
    h(r)=1-\frac{2 G M}{r}-\frac{(1-{\hat \alpha})}{24\pi G c_{\rm O}} r^2
\end{eqnarray}
is the lapse function following from (\ref{R0-gecici}). In the sequel, we will analyze this 
constant-curvature configuration ($R=R_0\neq 0$) in both the dS ($V_{\rm O}>0$ namely ${\hat \alpha}<1$) and AdS ($V_{\rm O}<0$ namely ${\hat \alpha}>1$) spacetimes in the case of rotating black holes.  We will use  rotating black hole properties to determine or constrain the model parameters $c_{\rm O}$ and ${\hat \alpha}$.

\section{Rotating Black holes in symmergent Gravity} \label{sec3}

Appropriate for a rotating black hole geometry, we use  Boyer-Lindquist coordinates with a metric free of the coordinate singularities in the spacetime both exterior to the black hole and interior to the cosmological horizon. The metric is given by \cite{Cembranos:2011sr,Perez:2012bx,Suvorov:2019qow,Myung:2013oca,Nashed:2021mpz,Nashed:2021sey}
\begin{equation}
\mathrm{d} s^{2}=\frac{\rho^{2}}{\Delta_{r}} \mathrm{~d} r^{2}+\frac{\rho^{2}}{\Delta_{\theta}} \mathrm{d} \theta^{2}+\frac{\Delta_{\theta} \sin ^{2} \theta}{\rho^{2}}\left[a \mathrm{d} t-\left(r^{2}+a^{2}\right) \frac{\mathrm{d} \phi}{\Xi}\right]^{2}-\frac{\Delta_{r}}{\rho^{2}}\left(\mathrm{d} t-a \sin ^{2} \theta \frac{\mathrm{d} \phi}{\Xi}\right)^{2}
\end{equation}
with the various quantities
\begin{eqnarray}
&&\Delta_{r}=\left(r^{2}+a^{2}\right)\left(1-\frac{r^2}{3}\frac{(1-\hat{\alpha})}{8 \pi c_{\rm O}}\right)-2 M r\,,\label{Deltar} \\
&&\rho^{2} =r^{2}+a^{2} \cos ^{2} \theta\,, \\
&&\Delta_{\theta} =1+\frac{ r^2}{3}\frac{(1-\hat{\alpha})}{8 \pi c_{\rm O}} a^{2} \cos ^{2} \theta\,, \\
&&\Xi =1+\frac{ r^2}{3}\frac{(1-\hat{\alpha})}{8 \pi c_{\rm O}} a^{2}\,.
\end{eqnarray}
Here, $a$ is the black hole spin parameter, and ${\hat \alpha}$ and $c_{\rm O}$ are the parameters of the symmergent gravity. The physical energy $E$ and angular momentum $J$ of the black hole are related to  the parameters $M$ and $a$ via the relations \cite{Gibbons:2004ai,Gao:2021xtt}
\begin{equation}
E=\frac{M}{\Xi^{2}}, \quad J=\frac{a M}{\Xi^{2}}
\label{EJ0}
\end{equation}
in which $M$ is the root of the equation $\Delta_{r}\left(r_{+}\right)=0$ namely
\begin{eqnarray}
M= \frac{1}{2 r_+}
\left(r_+^{2}+a^{2}\right)\left(1-\frac{r_+^2}{3}\frac{(1-\hat{\alpha})}{8 \pi c_{\rm O}}\right)
\end{eqnarray}
where $r_{+}$ is the radius of the event horizon. In this way, one can express energy and angular momentum in terms of the parameters $r_{+}$, $a$, and $G$. (One here notes that  $\frac{1}{l^2}={\frac{ (\hat{a}-1)}{24 \pi G\,c_{\rm O}}}$, where the curvature radius $l$ is related to the negative cosmological constant as $\Lambda= -3 l^{-2}$). As a result, one finds
\begin{equation}
  E=\frac{1}{2 \Xi^{2} r_{+} }\left(r_{+}^{2}+a^{2}+\frac{r_{+}^{4}}{l^{2}}+\frac{a^{2} r_{+}^{2}}{l^{2}}\right), \quad 
J=\frac{a}{2 \Xi^{2} r_{+} }\left(r_{+}^{2}+a^{2}+\frac{r_{+}^{4}}{l^{2}}+\frac{a^{2} r_{+}^{2}}{l^{2}}\right) 
\end{equation}
as the explicit expressions for the definitions in (\ref{EJ0}).

\begin{figure*}
   \centering
    \includegraphics[width=0.48\textwidth]{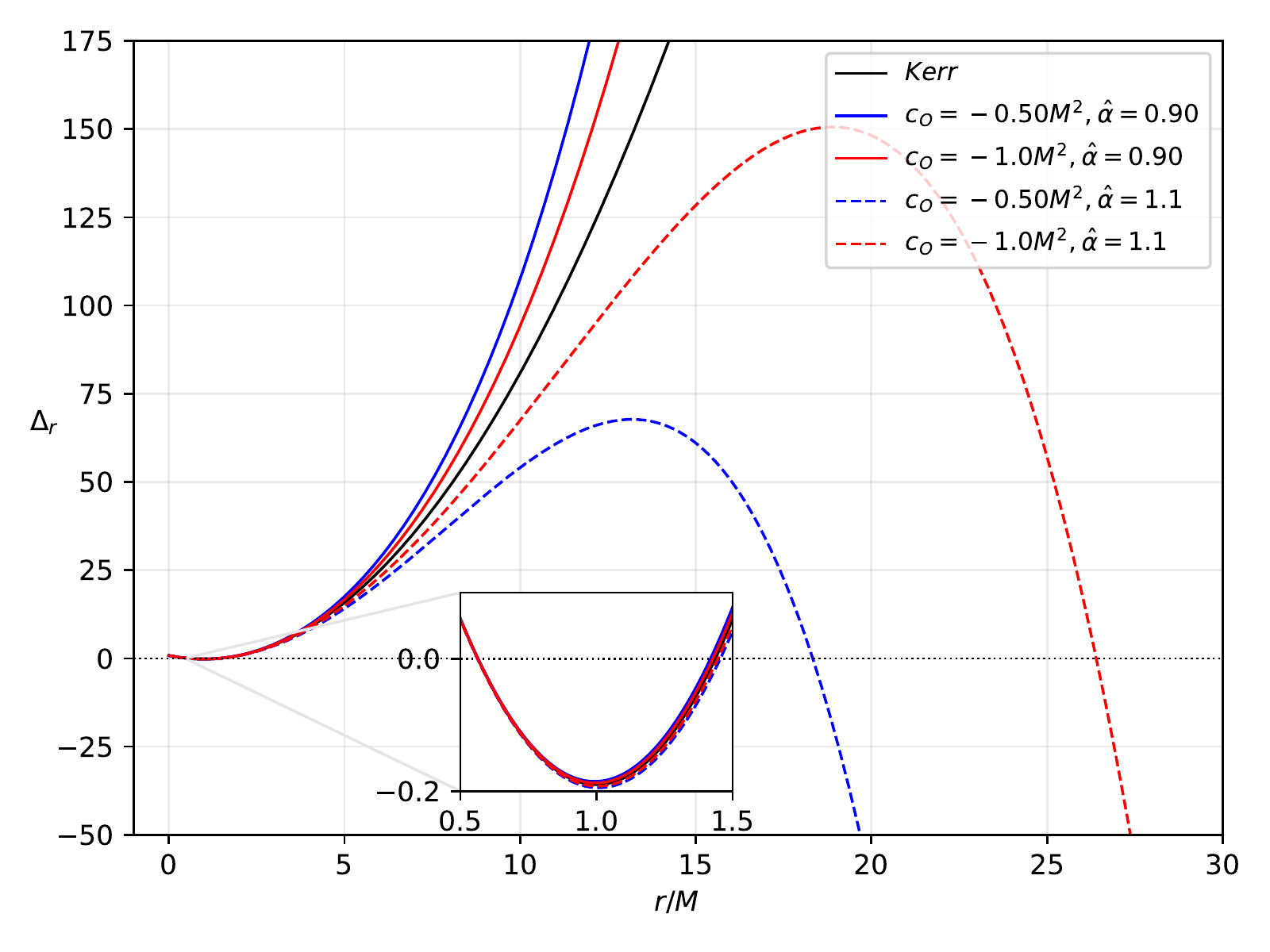} 
    \includegraphics[width=0.48\textwidth]{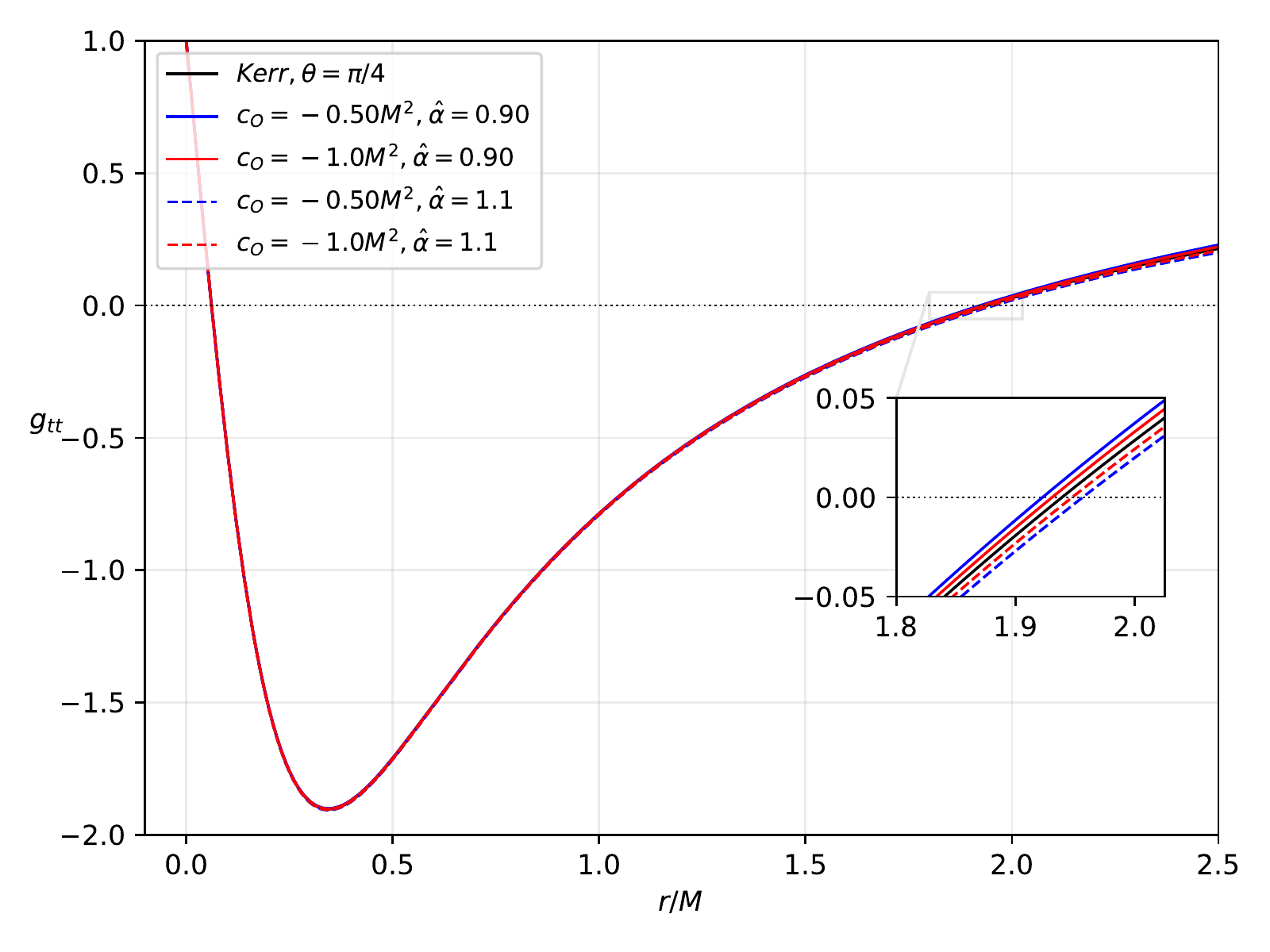}
    \caption{Variations of the functions $\Delta_r$ (left panel) and $g_{tt}$ (right panel) with the radial coordinate $r$ for the black hole spin parameter $a = 0.9\, M$. The plots show the null boundaries, with solid (dashed) lines mimicking the behavior of AdS (dS) black holes. We consider only negative values of $c_{\rm O}$ in agreement with \cite{Pogosian:2007sw}.}
    \label{hor}
\end{figure*}

Numerically, horizons can be determined by analyzing the lapse function, that is, by solving $\Delta_r = 0$ using its definition in (\ref{Deltar}). Besides, the location of the ergoregions can be determined numerically by plotting the metric component $g_{tt}$ and looking for points satisfying $g_{tt}=0$. In fact, plotted in Fig. \ref{hor} are $\Delta_r$ and $g_{tt}$ as functions of $r$ for the integration constant ${\hat \alpha}$ of $0.90$ and $1.10$,  black hole spin parameter $a = 0.9\, M$, and the various  values of $c_{\rm O}$. In general, symmergent black hole can mimic the dS (${\hat \alpha}>1$) or AdS-Kerr (${\hat \alpha}<1$) black holes depending on the sign of ${\hat \alpha}$. But, as already shown by Pogosian and Silvestri \cite{Pogosian:2007sw}, consistent f(R) gravity theories require $\frac{d F(R)}{d R}  > 0$ in order to remain stable the (non-tachyonic scalaron). This constraint imposes the condition $c_{\rm O} < 0$ on the loop-induced quadratic curvature coefficient, and restricts viable solutions to  the AdS-Kerr type. In view of the underlying QFT, $c_{\rm O} < 0$ implies that $n_\text{F} > n_\text{B}$ -- a mostly fermionic QFT \cite{demir1,demir2,demir3}. In the extreme case of no new bosons (in agreement with $n_\text{F} > n_\text{B}$), one finds out that the new particle sector remains inherently stable thanks to the fact that fermion masses do not have any power-law sensitivity to the UV cutoff.

Usually, value of the cosmological constant is scaled to a higher value to see its overall effect. In this sense,  assumed values of $\hat{\alpha}$ and $c_{\rm O}$ are also rescaled similarly.  As is seen from Fig. \ref{hor}, null boundaries are greatly affected by the values of the symmergent parameters $\hat{\alpha}$ and $c_{\rm O}$. Indeed, $\hat{\alpha}>1$ leads to three horizons where, in the context of the dS geometry, the third horizon corresponds to the cosmic outer boundary. It shifts  farther as $c_{\rm O}$ gets more negative. The two remaining null boundaries are the Cauchy and event horizons. In constrast to the dS solution, the symmergent AdS solution gives only two horizons such that symmergent effects are nearly vanishingly small on the inner horizon compared to the outer horizon. As for the ergoregions, $\hat{\alpha}>1$ leads to three ergoregions, and it can be seen that the farthest one turns out to be even beyond the cosmic horizon.

The Hawking temperature is given by
\begin{equation}
T=\frac{\kappa}{2\pi}
\end{equation}
such that the surface gravity 
\begin{equation}
\kappa^{2}=-\frac{1}{2}\triangledown^{\mu}\chi^{\nu}\triangledown_{\mu}\chi_{\nu}
\end{equation}
involves the null Killing vectors $\chi^{\nu}$. The two
Killing vectors $\xi^{\nu}=\partial_{t}$ and $\zeta^{\nu}=\partial_{\varphi}$ in the metric are associated with the time translation and rotational invariance,
respectively. Consequently, we take
\begin{equation}
\chi^{\nu}=\xi^{\nu}+\Omega\zeta^{\nu}
\end{equation}
and determine $\Omega$ under the condition that $\chi^{\nu}$ is a null vector. This leads to the constraint 
\begin{equation}
\chi^{\nu}\chi_{\nu}=g_{tt}+2\Omega g_{t\varphi}+\Omega^{2}g_{\varphi\varphi}=0,
\end{equation}
from which one gets
\begin{equation}
\Omega=-\frac{g_{t\varphi}}{g_{\varphi\varphi}}\pm\sqrt{\left(\frac{g_{t\varphi}}{g_{\varphi\varphi}}\right)^{2}-\frac{g_{tt}}{g_{\varphi\varphi}}}
\end{equation}
which reduces to
\begin{equation}
\Omega_{+}=\frac{a\Xi}{r_{+}^{2}+a^{2}} \label{eq:omega}
\end{equation}
at the event horizon $\Delta\left(r_{+}\right)=0$. For this $\Omega$ value one then gets
\begin{equation}
\kappa=\frac{1}{2\left(r_{+}^{2}+a^{2}\right)}\left.\frac{d\Delta_{r}}{dr}\right|_{r=r_{+}},
\end{equation}
for the surface gravity, 
\begin{equation}
    T=\frac{r_{+}}{4 \pi\left(r_{+}^{2}+a^{2}\right)}\left(1+\frac{a^{2}}{l^{2}}+\frac{3 r_{+}^{2}}{l^{2}}-\frac{a^{2}}{r_{+}^{2}}\right)
=\frac{3 r_{+}^{4}+\left(a^{2}+l^{2}\right) r_{+}^{2}-l^{2} a^{2}}{4 \pi l^{2} r_{+}\left(r_{+}^{2}+a^{2}\right)}\label{hawking}
\end{equation}
for the Hawking temperature, and
\begin{equation}
    S=\frac{A\left[1+f'(R_0)\right]}{4}=\frac{\pi\left(r_{+}^{2}+a^{2}\right)}{ \Xi}
\end{equation}
for the Bekenstein-Hawking entropy. Needless to say,  all thermodynamic quantities are necessarily non-negative. 

The rotating symmergent black hole satisfies therefore the  first law of thermodynamics 
\begin{equation}
d E=T d S+\Omega d J
\end{equation}
with 
\begin{eqnarray}
\Omega=\frac{a\left(1+r_{+}^{2} l^{-2}\right)}{r_{+}^{2}+a^{2}}\,.
\end{eqnarray}

\section{Geodesics around rotating symmergent black holes} \label{sec4}
In this section we give a detailed analysis of the geodesics and orbits around the symmergent black holes.
\subsection{Null geodesic and shadow cast}
We start the analysis with null geodesics. The Hamilton-Jacobi equation gives  
\begin{equation} \label{e33}
    \frac{\partial S}{\partial \lambda}=-H,
\end{equation}
where $S$ is the Jacobi action in terms of the affine parameter $\lambda$ (proper time) and coordinates $x^{\mu }$. The Hamiltonian is given by
\begin{equation} \label{e34}
    H=\frac{1}{2}g^{\mu \nu }\frac{
    \partial S}{\partial x^{\mu }}\frac{\partial S}{\partial x^{\nu }},
\end{equation}
in the GR so that 
\begin{equation} \label{e35}
    \frac{\partial S}{\partial \lambda }=-\frac{1}{2}g^{\mu \nu }\frac{
    \partial S}{\partial x^{\mu }}\frac{\partial S}{\partial x^{\nu }}
\end{equation}
as follows from \eqref{e33} above. Using the separability ansatz for the Jacobi function 
\begin{equation} \label{e36}
    S=\frac{1}{2}\mu ^{2}\lambda -Et+L\phi +S_{r}(r)+S_{\theta }(\theta),
\end{equation}
with the particle mass $\mu$, one is led to the following first-order motion equations \cite{Slany2020}
\begin{align} \label{eos}
    &\Sigma\frac{dt}{d\lambda}=\frac{\Xi (r^2+a^2)P(r)}{\Delta _r}-\frac{\Xi aP(\theta )}{\Delta _{\theta }}, \nonumber \\
    &\Sigma\frac{dr}{d\lambda}=\sqrt{R(r)}, \nonumber \\
    &\Sigma\frac{d\theta}{d\lambda}=\sqrt{\Theta(\theta)}, \nonumber \\
    &\Sigma\frac{d\phi}{d\lambda}=\frac{\Xi aP(r)}{\Delta _r}-\frac{\Xi P(\theta )}{\Delta _{\theta }\sin ^2\theta }
\end{align}
after introducing the functions
\begin{align}
    &R(r) = P(r)^2 - \Delta _r(\mu^2r^2+K), \nonumber \\
    &P(r) = \Xi E(r^2+a^2)-\Xi aL, \nonumber \\
    &\Theta(\theta) = \Delta_{\theta }(K-\mu^2a^2\cos ^2\theta ) - \frac{P(\theta )^2}{\sin^2\theta }, \nonumber \\
    & P(\theta) = \Xi(aE\sin ^2\theta -L ).
\end{align}
From the third equation in \eqref{eos} above, constants of motion can be correlated via the relation $K = \Xi^2(aE-L)^2$, which is a consequence of a hidden symmetry in the $\theta$-coordinate \cite{Slany2020,carter1968global}.

\begin{figure*}
   \centering
    \includegraphics[width=0.48\textwidth]{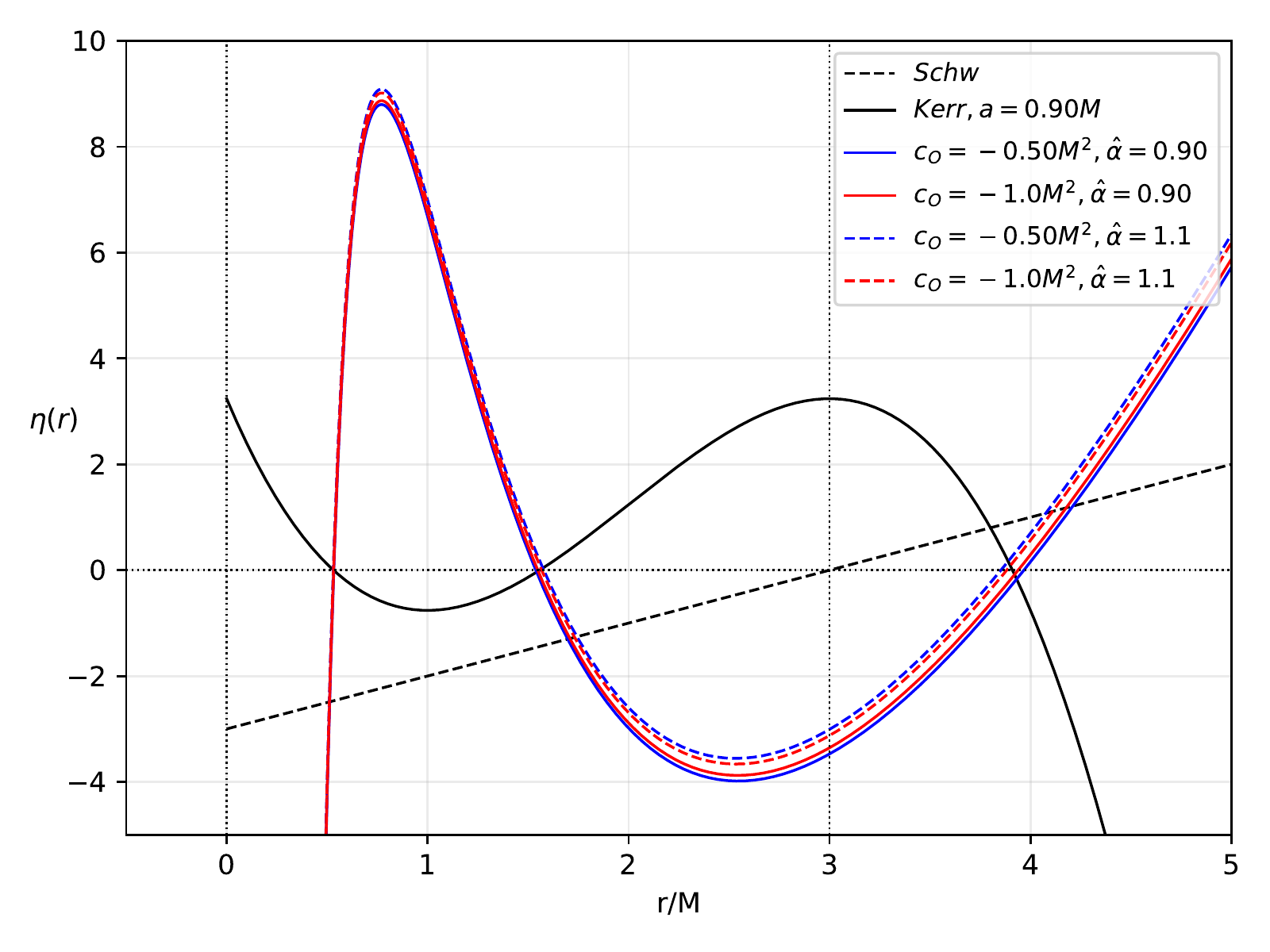} 
    \includegraphics[width=0.48\textwidth]{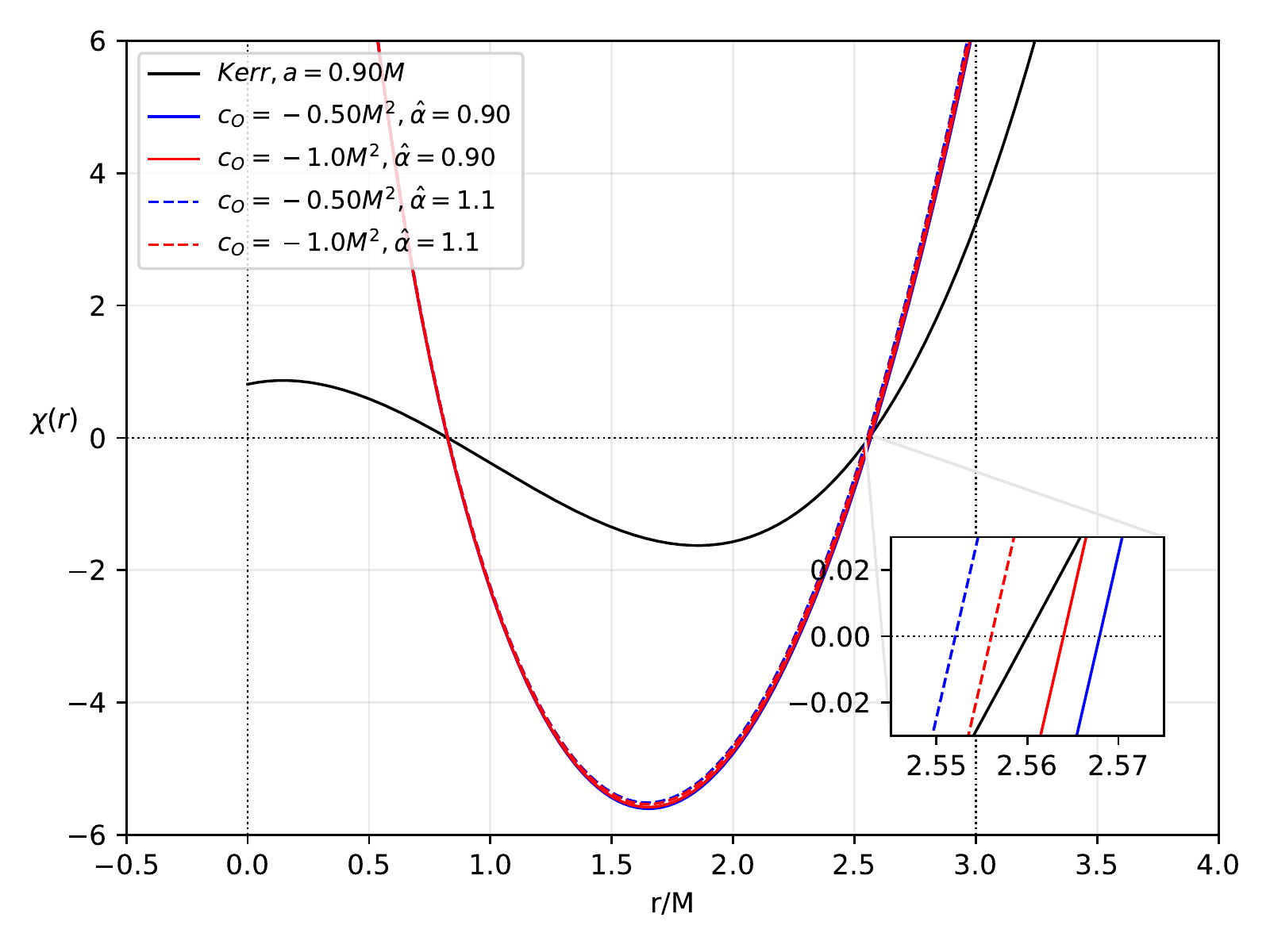}
    \caption{Location of the photonsphere (left panel), and location of the photonsphere crossing at $\phi=0$ or along the equatorial plane (right panel). We take $a=0.90M$ in both plots ($a$ is the black hole spin parameter).}
    \label{rph}
\end{figure*}

First, we study null geodesic and shadow cast for massless particles ($\mu=0$). Photon circular orbits, which are always unstable, must then satisfy the following condition
\begin{equation} \label{e39}
    R(r)=\frac{dR(r)}{dr}\mid_{r=r_{\text{o}}}=0.
\end{equation}
The null geodesic is important is an important quantity for symmergent black holes.  We need, in particular, the photon region in constant $r_0$, the so-called photonsphere, which is an unstable orbit. The shadow cast ultimately depends on the photon region. With $\mu=0$, it proves convenient to define these two impact parameters
\begin{equation} \label{e45}
    \xi=\frac{L}{E} \quad {\rm and}\quad \eta=\frac{K}{E^2}.
\end{equation}
These parameters are found to possess the following explicit expressions
\begin{eqnarray} 
    \xi&=&\frac{\Delta_r'(r^{2}+a^{2})-4\Delta_r r}{a\Delta_r'},\label{e46}\\
    \eta&=&\frac{-r^{4}\Delta_r'^{2}+8r^{3}\Delta_r\Delta_r'+16r^{2}\Delta_r(a^{2}-\Delta_r)}{a^{2}\Delta_r'^{2}}\label{e47}
\end{eqnarray}
after a lengthy algebra. The photonsphere radius $r_\text{ph}$ can then be determined by solving $\eta(r)=0$ for $r$. The analytical solutions are well-known for both Schwarzschild and Kerr black holes. In our case, we plot \eqref{e47} in Fig. \ref{rph} to get the qualitative impression about the location of $r_\text{ph}$.
The left panel of Fig. \ref{rph} concerns location of the photonsphere along the equatorial plane. As we would expect, the Schwarzschild case gives $r_\text{ph}=3M$. The extreme Kerr case gives two possible locations, that is, the prograde orbit  at $r_\text{ph}=M$ and the retrograde orbit at $r_\text{ph}=4M$. The plot shows the locations for $a=0.90M$, which fall near the said extreme values. For the symmergent effect, the dS type (dashed lines) gives slightly lower values in the retrograde case while the AdS type gives a larger one. It is clear that the more negative the $c_{\rm O}$ the closer the photon ring to the Kerr case. The symmergent effect is also present in the prograde case though the deviation is smaller than in the retrograde case. In the right panel of Fig. \ref{rph}, we consider photons with zero angular momentum, that is, those that traverse the equatorial plane in a perpendicular manner (the so-called nodes). The retrograde case gives higher values for such an orbit than the prograde case. The deviation caused is barely evident, especially in the prograde case (the inset highlights the deviation). As a final remark, we note that the symmergent effect mimics dS or AdS cases, where such a behavior does not occur in the Schwarzschild case. The coupling between the spin and symmergent parameters made it possible to affecy the behavior of the photonsphere.

\begin{figure*}
   \centering
    \includegraphics[width=0.48\textwidth]{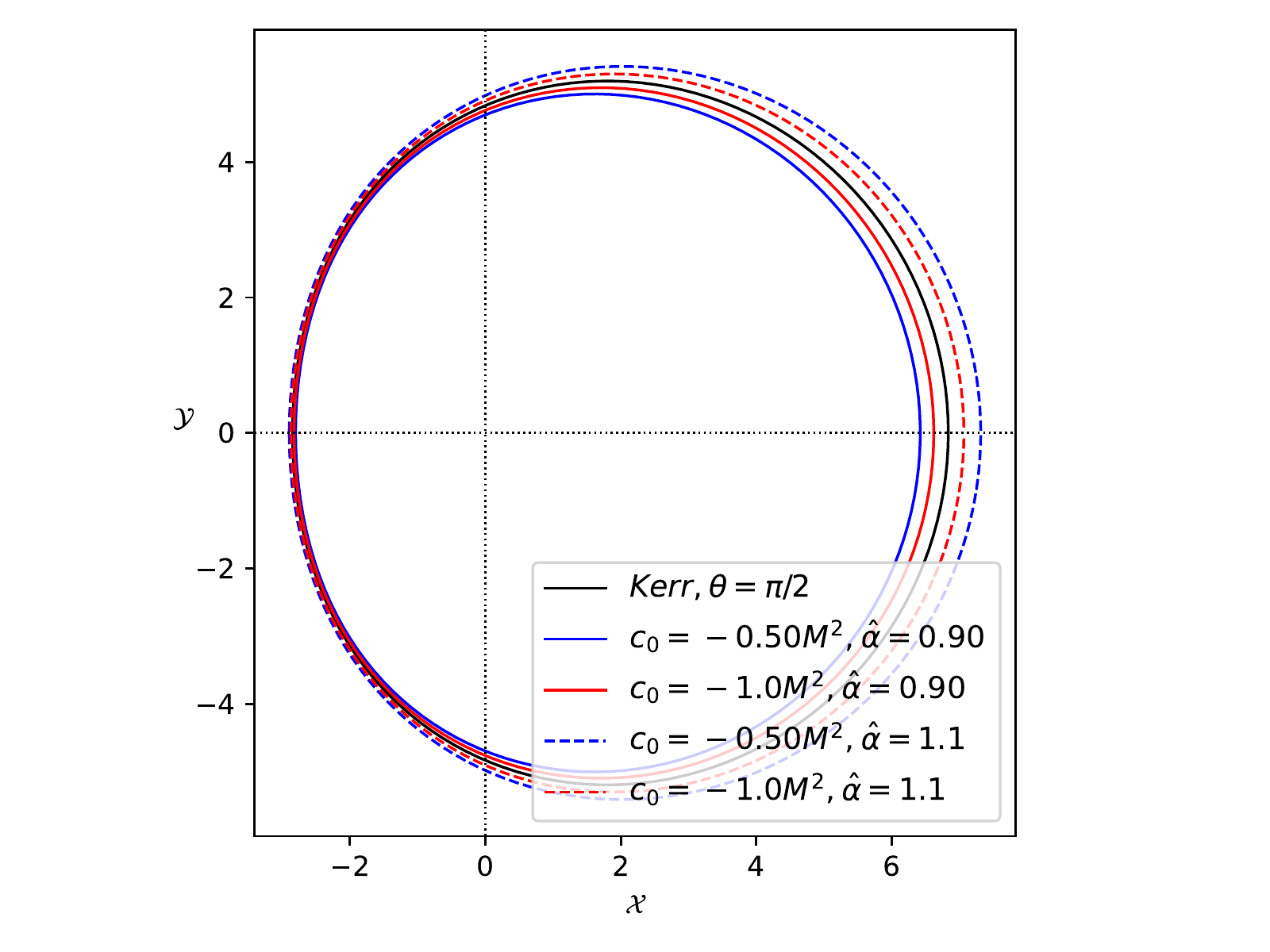} 
    \includegraphics[width=0.48\textwidth]{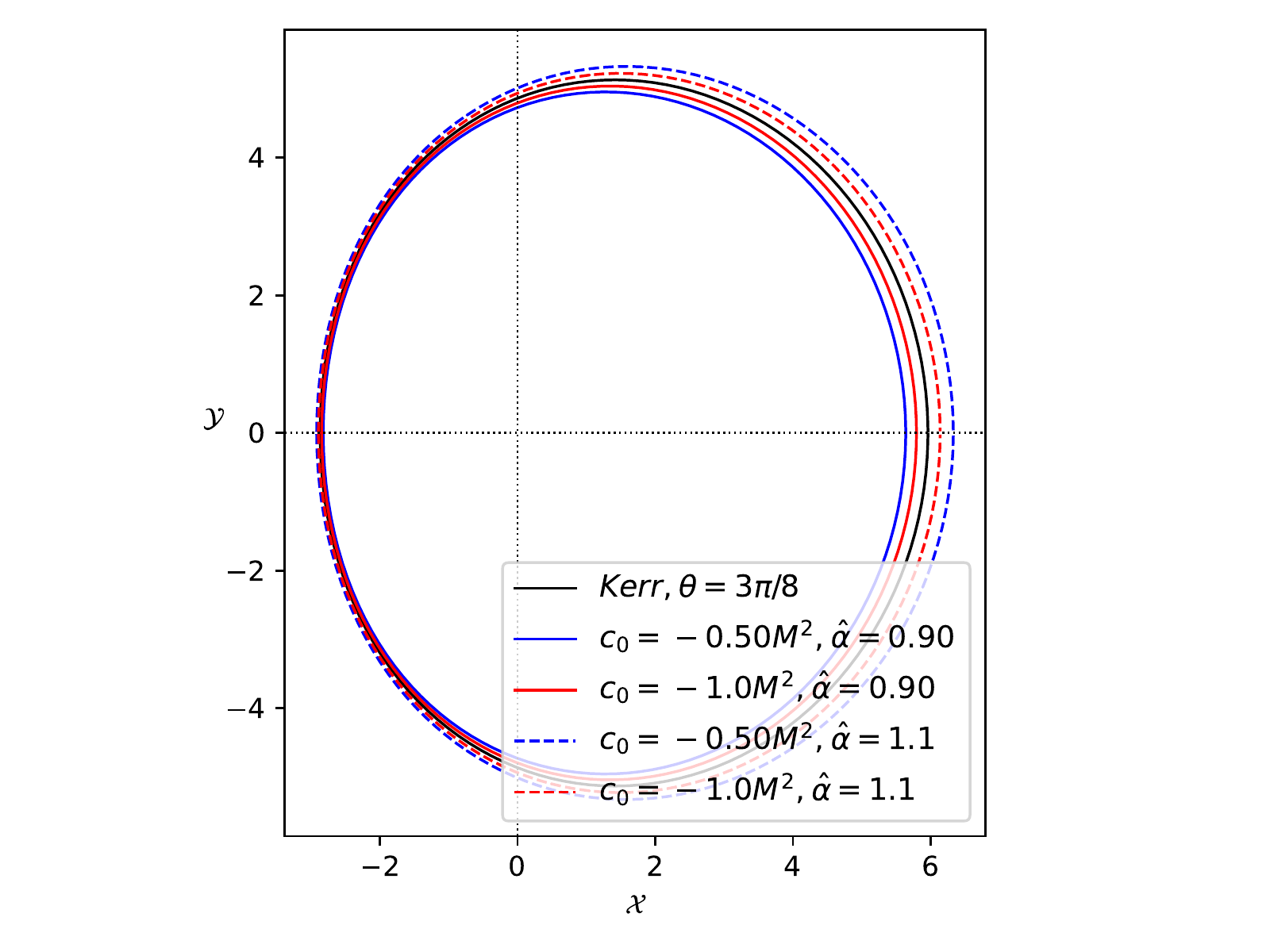}
    \includegraphics[width=0.48\textwidth]{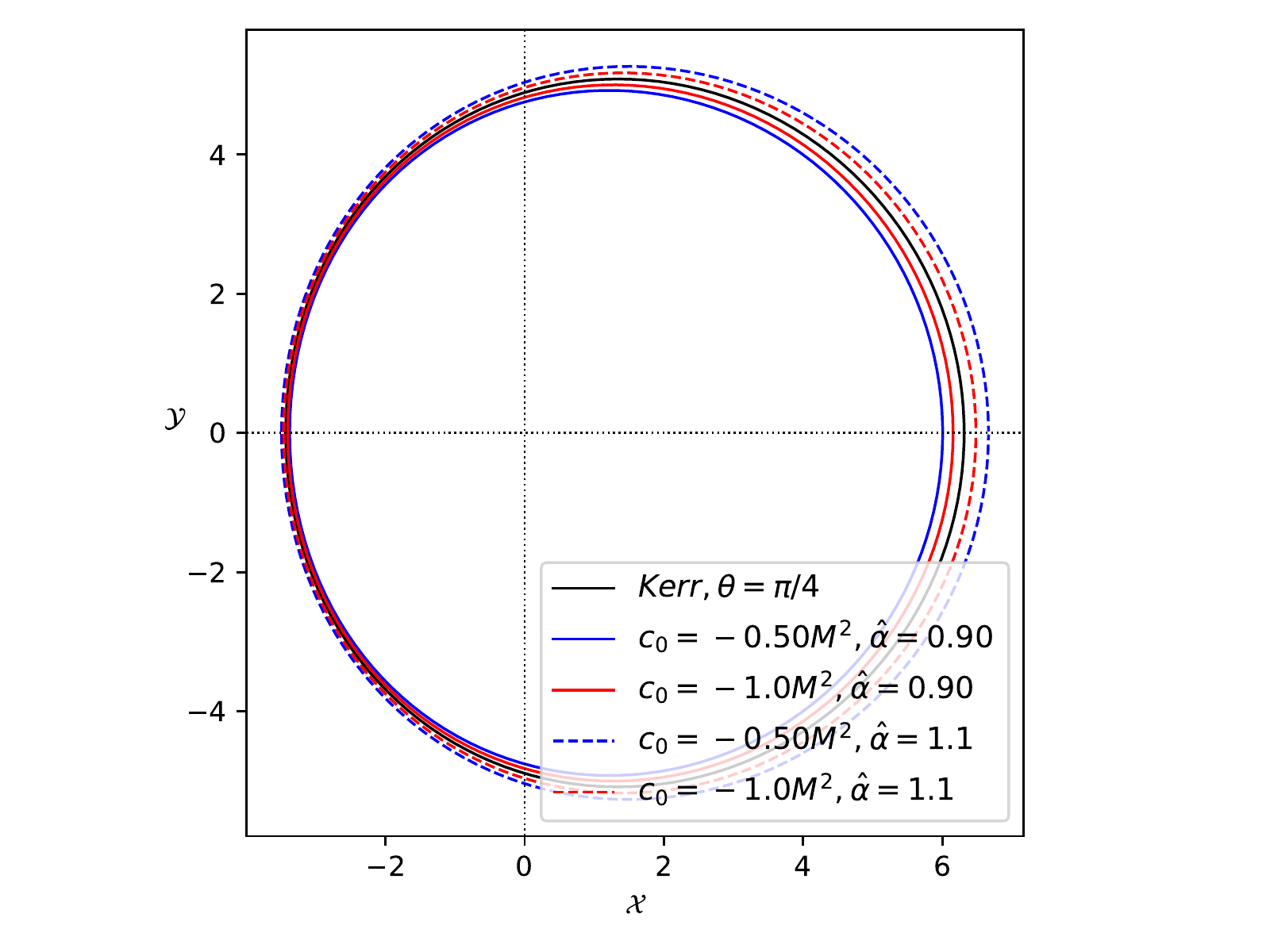}
    \includegraphics[width=0.48\textwidth]{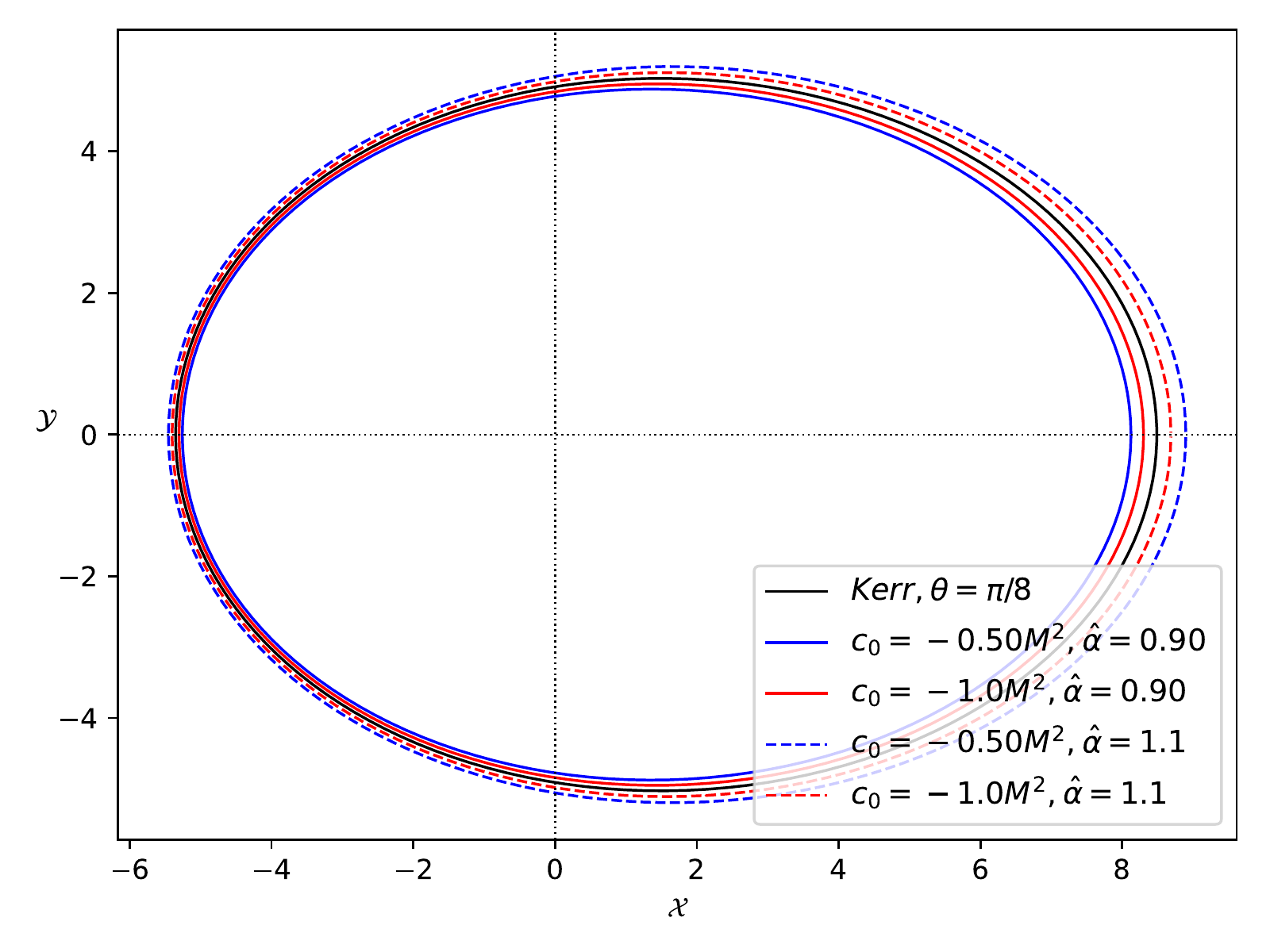}
    \caption{Shadow plot for different inclination angles for  the black hole spin parameter $a=0.90M$ (aspect ratios of figures are set to obtain the actual view).}
    \label{shad}
\end{figure*}

Escaping photons in the unstable orbit gives the possibility for remote observers to backward-trace and obtain a shadow cast by using the celestial coordinates $(r_{\text{obs}},\theta_{\text{obs}})$. Such an observer is also known as the Zero Angular Momentum Observer (ZAMO). Also, in this type of co-moving frame, without loss of generality, one makes the approximation $r_{\text{obs}} \rightarrow \infty$  and takes $\theta_{\text{obs}} = \pi/2$. The  celestial coordinates are defined as \cite{Johannsen2013}
\begin{align} \label{e48}
    \mathcal{X}  &=-r_{\text{obs}}\frac{\xi }{\zeta\sqrt{g_{\phi \phi }} \left( 1+\frac{g_{t\phi }}{g_{\phi \phi }}\xi\right) }, \nonumber\\
    \mathcal{Y}  &= r_{\text{obs}}\frac{\pm \sqrt{\Theta (i)}}{\zeta\sqrt{g_{\theta \theta }} \left( 1+\frac{g_{t\phi}}{g_{\phi \phi }}\xi \right) },
\end{align}
and the condition $r_{\text{obs}} \rightarrow \infty$  leads to the simplified relations
\begin{align} \label{e49}
    \mathcal{X}&=-\xi \csc \theta_{\text{obs}},   \nonumber \\
    \mathcal{Y}&=\pm \sqrt{\eta +a^{2}\cos ^{2}\theta_{\text{obs}}-\xi ^{2}\cot^{2}\theta_{\text{obs}}}.
\end{align}
These expressions further simplify to $\mathcal{X} = 0$ and $\mathcal{Y}= \pm \sqrt{\eta}$ when $\theta_{\text{obs}}=\pi/2$. If, furthermore, $a=0$ then shadow cast of a Schwarzschild black hole (a circle) is obtained. The plot of $\mathcal{Y}$ vs. $\mathcal{X}$ is shown in Fig. \ref{shad} for the black hole spin parameter value of  $a=0.90M$.
Overall, we observe the D-shaped nature of the shadow cast due to the spin parameter. Considering first  the effect of the inclination angle, we see that as we decrease the value of $\theta$ (going up from the equatorial plane), we see that the shape changes. Remarkably, in the upper right figure, the shape becomes more compressed in the $\alpha$ axis, and as one continues to go up (lower left figure), it goes back to an almost spherical shape. Finally, near the pole, a drastic change in the shadow shape occurs due to stretching of the $\alpha$-axis. For the symmergent effects, we note that the dS type tends to decrease the photonsphere radius. Nevertheless, as photons travel in the intervening space under the effect of the symmergent parameter the shadow size is seen to increase. We can see the contrast in the AdS type symmergent effect in that while the photonsphere increases the shadow size tends to decrease.

\begin{figure}
   \centering
    \includegraphics[width=0.48\textwidth]{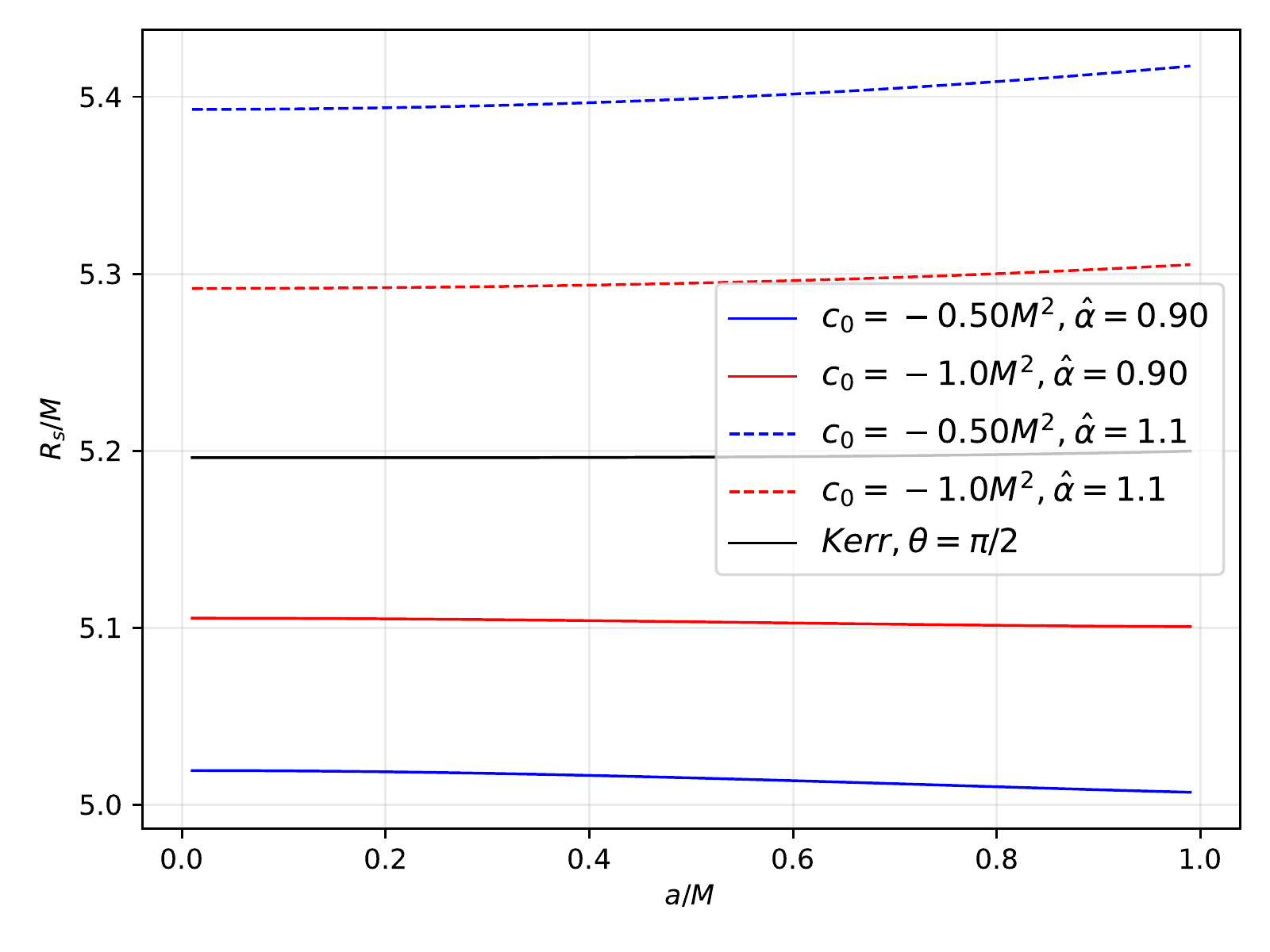}
    \includegraphics[width=0.48\textwidth]{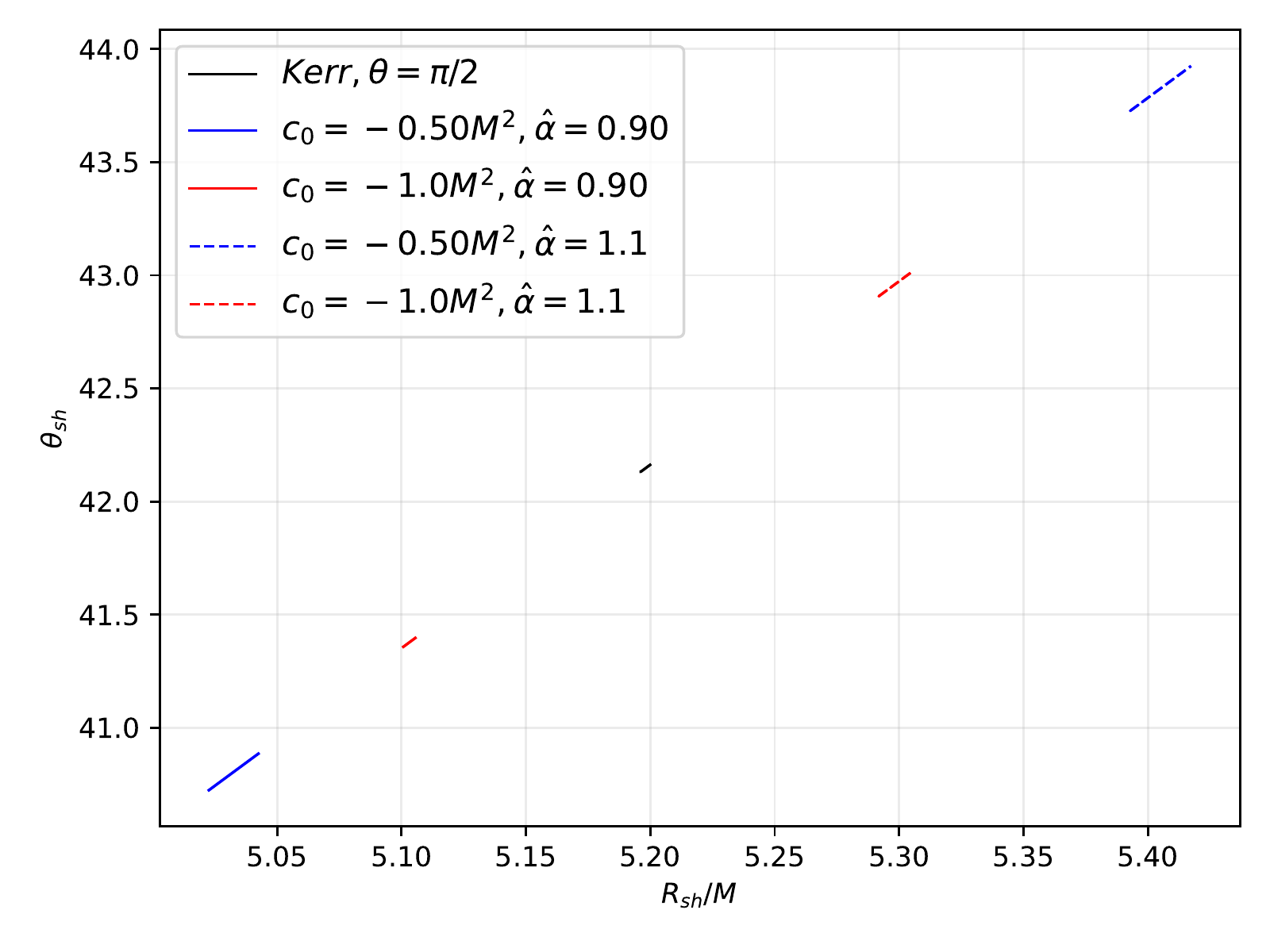}
    \includegraphics[width=0.48\textwidth]{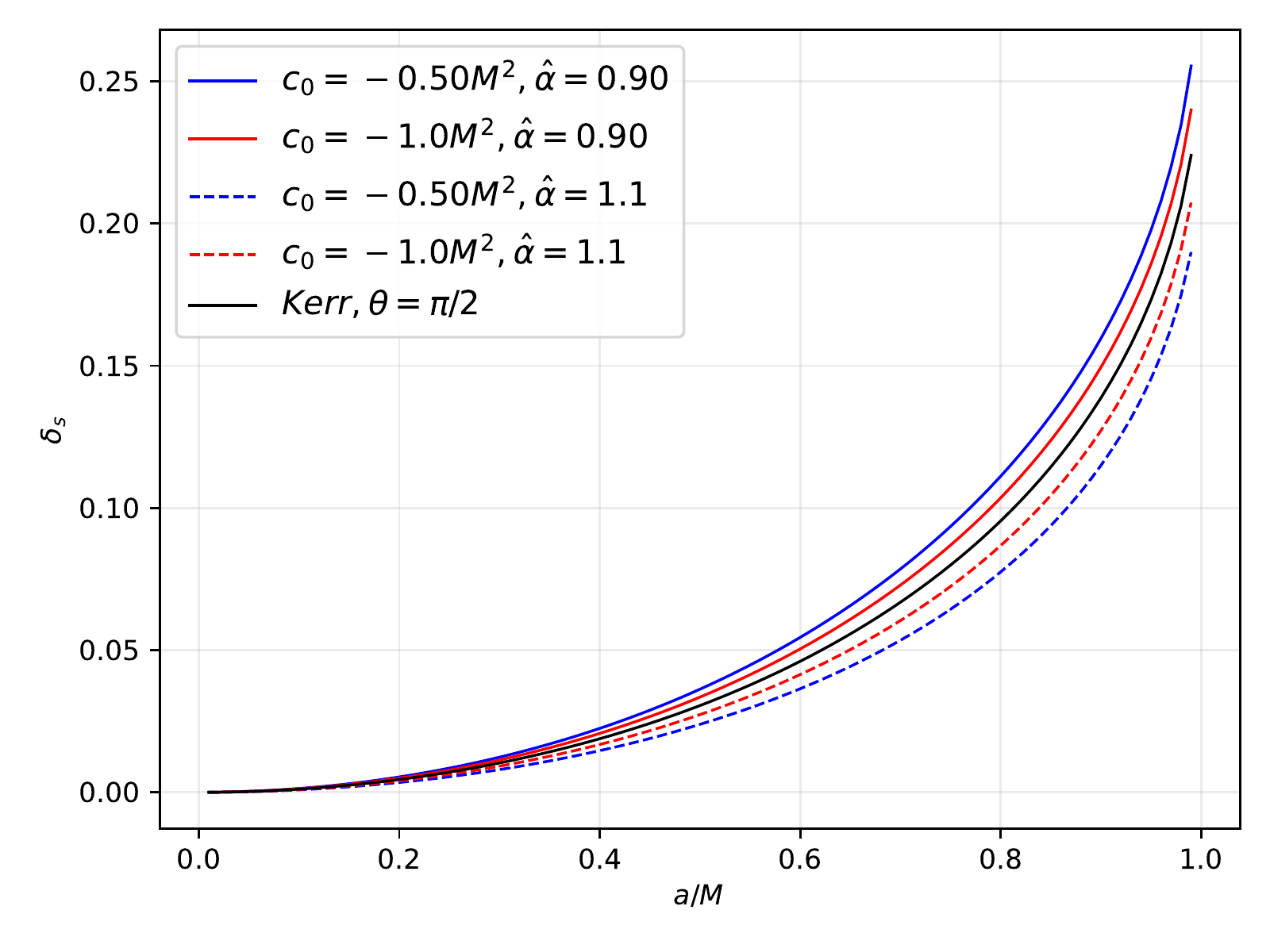}
    \includegraphics[width=0.48\textwidth]{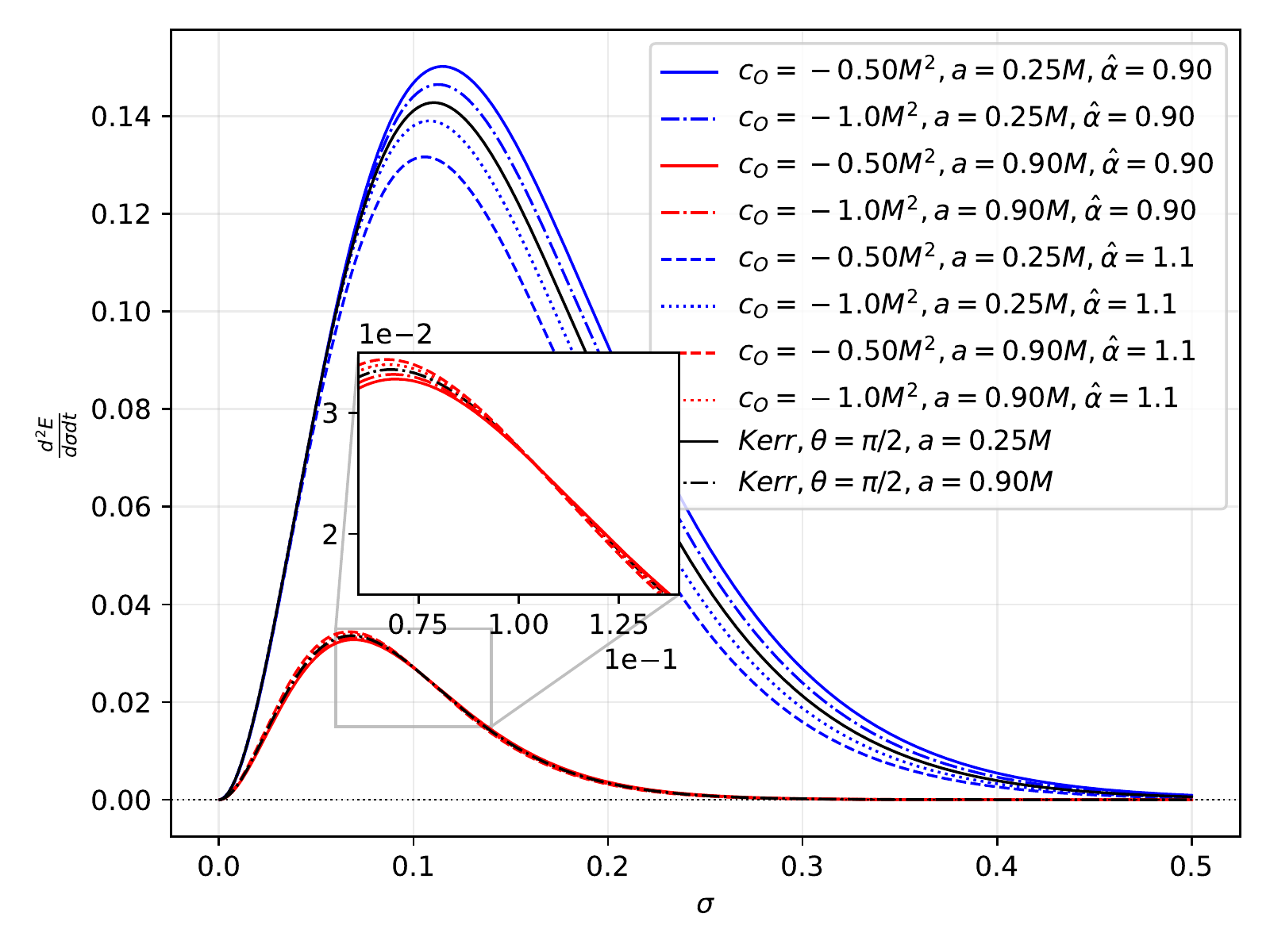}
    \caption{Shadow observables associated with the upper left panel of Fig. \ref{shad}. Shown here are shadow radius (upper-left panel), angular diameter in $\mu$as for M87* (upper-right panel), distortion parameter (lower-left panel), and energy emission rate (lower-right panel).}
    \label{shaobs}
\end{figure}

In the azimuthal plane, we see how the shadow becomes ``D-shaped" when the balck hole spin parameter $a$ is near extremal. The numerical value of the shadow radius associated with this shape \cite{Hioki2009} can be calculated \cite{Dymnikova2019} by using
\begin{equation} \label{eq-rad}
    R_\text{s}=\frac{\mathcal{Y}_{\text{t}}^2+(\mathcal{X}_{\text{t}}-\mathcal{X}_{\text{r}})^2}{2|\mathcal{X}_{\text{t}}-\mathcal{X}_{\text{r}}|}
\end{equation}
which we plot in Fig. \ref{shaobs} (upper left), along with the shadow's angular radius $\theta_{\text{sh}}$ (upper right), where the latter is defined by
\begin{equation} \label{eq-ang}
    \theta_{\text{sh}}=9.87098\times10^{-3} \frac{R_{\text{s}}M}{D}
\end{equation}
after taking the black hole mass  $M$ in units of $M_{\odot}$ and $D$ in parsecs. We emphasize that these behaviors are consistent with Fig. \ref{shad}. Other observables that can be derived from the shadow are the distortion parameter $\delta_\text{s}$ and the energy emission rate $\frac{d^{2}E}{d\omega dt}$, which are defined as
\begin{equation} \label{eq-dis}
    \delta_\text{s}=\frac{d_{\text{s}}}{R_{\text{s}}}=\frac{\tilde{\mathcal{X}}_{\text{l}}-\mathcal{X}_{\text{l}}}{R_{\text{s}}},
\end{equation}
\begin{equation} \label{eq-eer}
\frac{d^{2}E}{d\omega dt}=2\pi^{2}\frac{\Pi_{ilm}}{e^{\omega/T}-1}\omega^{3}
\end{equation}
such that the energy absorption cross-section can be approximated as $\Pi_{ilm} \sim \pi R_{\text{s}}^2$  for an observer at $r_\text{obs} \rightarrow \infty$. These observables are plotted in  Fig. \ref{shaobs} in the lower two panels. From these panels we are able to see how the spin parameter increases the distortion of the shadow. We can also see that even if the dS symmergent effect tends to increase the shadow, the distortion it gives is smaller than the AdS symmergent effect.
In the lower right panel of Fig. \ref{shaobs}, we comparatively show $a=0.25M$ and $a=0.90M$ for the Kerr, dS and AdS symmergent black holes. For the Kerr case, we see that low values of $a$ leads to higher values for energy emission rate and peak frequency. Adding the symmergent effect at low values of $a$ and $\hat{\alpha} = 0.90$, we see that a less negative $c_{\rm O}$ provides the highest peak frequency and energy emission rate. For $\hat{\alpha} = 1.1$, this very same $c_{\rm O}$ gives the lowest emission rate and peak frequency. For $a = 0.90M$, we observe in the inset plot that the roles are reversed near the peak of the curve. We see that there is a point in the plot where the behavior flips as the frequency $\sigma$ increases whilst the emission rate decreases.


We now investigate observational constraints on the symmergent parameter $c_{\rm O}$ using the black hole shadow data collected from M87* \cite{EventHorizonTelescope:2019dse} and Sgr. A* \cite{EventHorizonTelescope:2022xnr}. The data is tabulated in Table \ref{tab_obs}.
\begin{table}
    \centering
    \begin{tabular}{ p{2cm} p{3cm} p{4cm} p{2cm}}
    \hline
    \hline
    Black hole & Mass ($M_\odot$) & Angular diameter: $2\theta_\text{sh}$ ($\mu$as) & Distance (kpc) \\
    \hline
    Sgr. A*   & $4.3 \pm 0.013$x$10^6$ (VLTI)    & $48.7 \pm 7$ (EHT) &   $8.277 \pm 0.033$ \\
    M87* &   $6.5 \pm 0.90$x$10^9$  & $42 \pm 3$   & $16800$ \\
    \hline
    \end{tabular}
    \caption{Black hole observational constraints.}
    \label{tab_obs}
\end{table}
With these data, one can determine  diameter of the shadow size in units of the black hole mass with
\begin{equation} \label{ed}
    d_\text{sh} = \frac{D \theta}{M}.
\end{equation}
In this sense, the diameter of the shadow image of M87* and Sgr. A* are $d^\text{M87*}_\text{sh} = (11 \pm 1.5)m$ and $d^\text{Sgr. A*}_\text{sh} = (9.5 \pm 1.4)m$, respectively. Meanwhile, the theoretical shadow diameter can be obtained via $d_\text{sh}^\text{theo} = 2 R_\text{sh}$ with the use of equation \eqref{eq-rad}.
\begin{figure}
   \centering
    \includegraphics[width=0.48\textwidth]{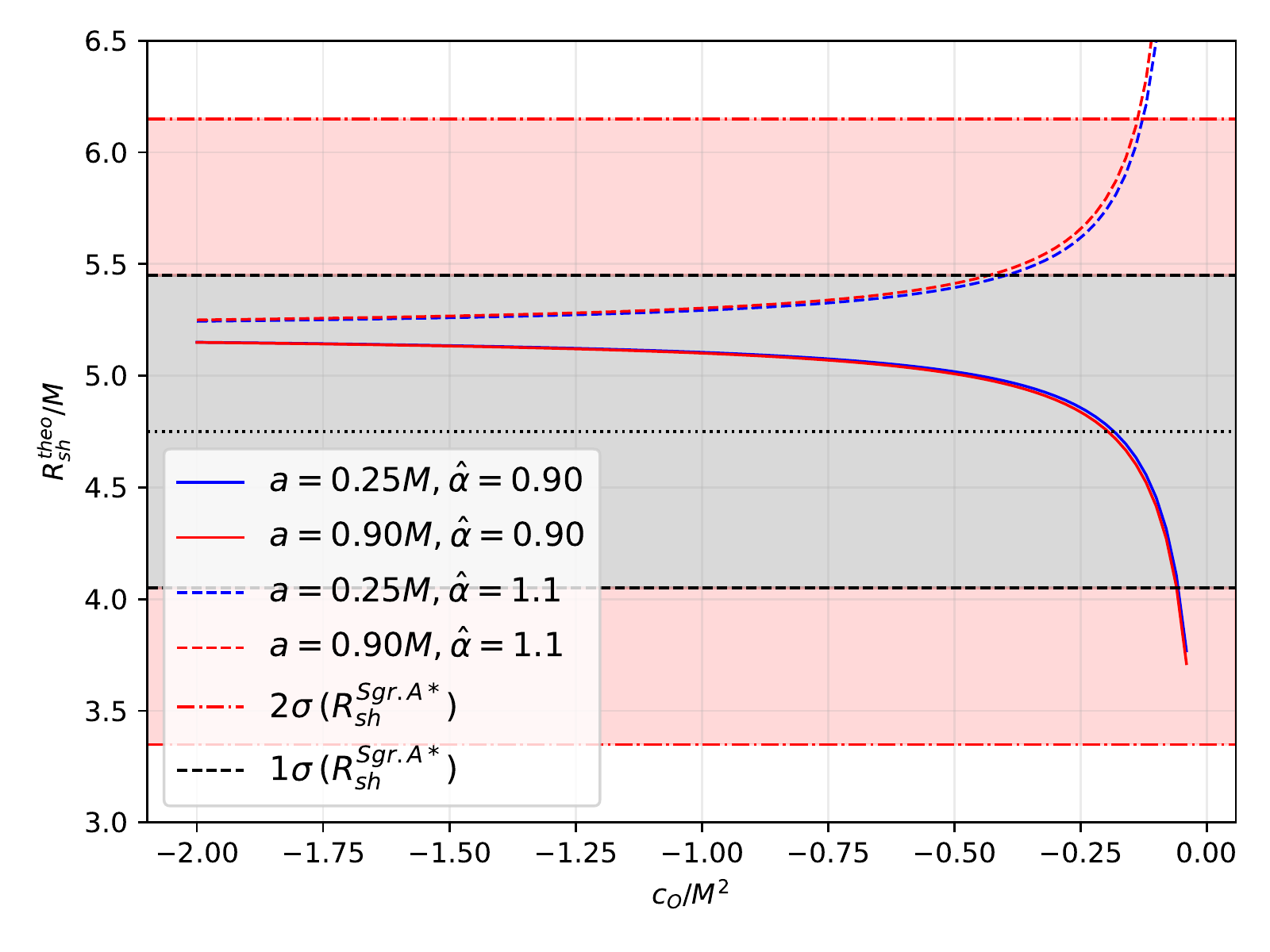}
    \includegraphics[width=0.48\textwidth]{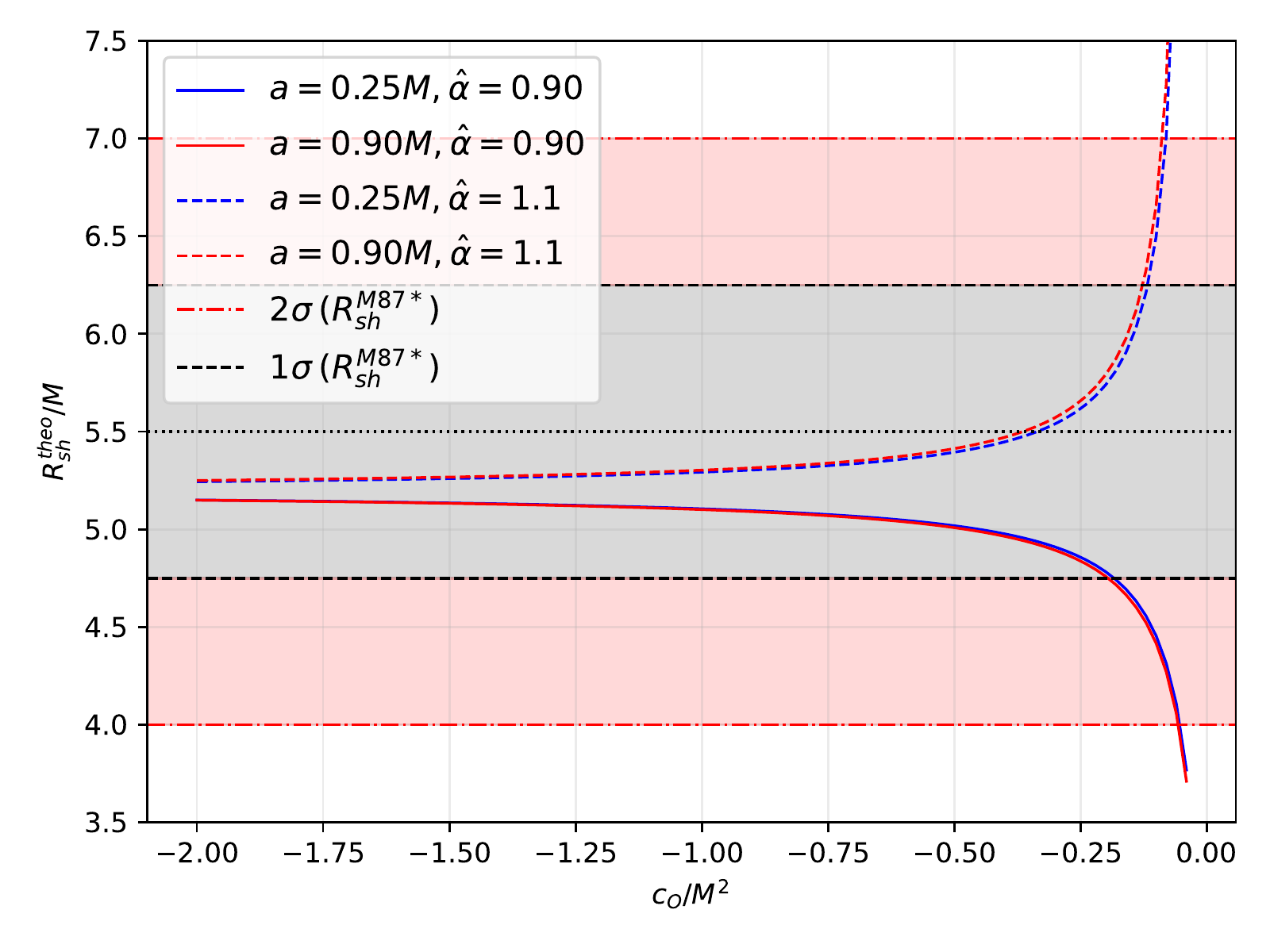}
    \caption{Constraint on the symmergent gravity parameter $c_{\rm O}$ by the Sgr. A* data (left-panel) and M87* data (right-panel).}
    \label{sha_cons}
\end{figure}
Based on Fig. \ref{sha_cons}, we can conclude that the parameters we used in describing the shadow cast fall within the upper and lower bounds at the $1\sigma$ level \cite{EventHorizonTelescope:2019dse, EventHorizonTelescope:2022xnr,EventHorizonTelescope:2021dqv,Vagnozzi:2022moj}. Where this happens, the range in $c_{\rm O}$ is more extensive in M87* than in Sgr. A*. We also observe that less negative $c_{\rm O}$ introduces more sensitivity to deviation than more negative $c_{\rm O}$.
\begin{table}[!ht]
    \centering
    \begin{tabular}{ l|cc|cc|c }
    \hline
    \hline
      Sgr. A* &  \multicolumn{2}{c|}{$2\sigma$} & \multicolumn{2}{c|}{$1\sigma$} & \multicolumn{1}{c}{observed $R_\text{sh}$} \\
    \hline
    spin $a$ (${\hat \alpha = 0.90}$) & upper & lower & upper & lower  & mean   \\
    \hline
    $0.25M$ & - & - & - & -0.057  & -0.184    \\
    $0.90M$ & - & - & - & -0.060  & -0.196   \\
    \hline
    spin $a$ (${\hat \alpha = 1.10}$) & upper & lower & upper & lower  & mean  \\
    \hline
    $0.25M$ & -0.127 & - & -0.397 & - & - \\
    $0.90M$ & -0.137 & - & -0.431 & - & - \\
    \hline
    \end{tabular}
    \quad
    \begin{tabular}{ l|cc|cc|c }
    \hline
    \hline
      M87* &  \multicolumn{2}{c|}{$2\sigma$} & \multicolumn{2}{c|}{$1\sigma$} & \multicolumn{1}{c}{observed $R_\text{sh}$} \\
    \hline
    spin $a$ (${\hat \alpha = 0.90}$) & upper & lower & upper & lower  & mean   \\
    \hline
    $0.25M$ & - & -0.054 & - & -0.196  & -    \\
    $0.90M$ & - & -0.057 & - & -0.185  & -   \\
    \hline
    spin $a$ (${\hat \alpha = 1.10}$) & upper & lower & upper & lower  & mean  \\
    \hline
    $0.25M$ & -0.081 & - & -0.117 & - & -0.336 \\
    $0.90M$ & -0.089 & - & -0.128 & - & -0.364 \\
    \hline
    \end{tabular}
    \caption{Bounds on $c_{\rm 0}$ as read off from the curves in Fig. \ref{sha_cons}.}
    \label{Tab2}
\end{table}

\subsection{Symmergent gravity effects on time-like orbits: Effective potential and ISCO radii}
We now turn our attention to time-like orbits. Here, we will still use the equations of motion from the Hamilton-Jacobi approach but this time set $\mu=1$. We are interested in determining the radius of the innermost stable circular orbit (ISCO), which is very important in the study of matter accretion disks and in the study also of the effective potential for getting a qualitative description of bound and unbound orbits under the symmergent effects.

\begin{figure*}
   \centering
    \includegraphics[width=0.48\textwidth]{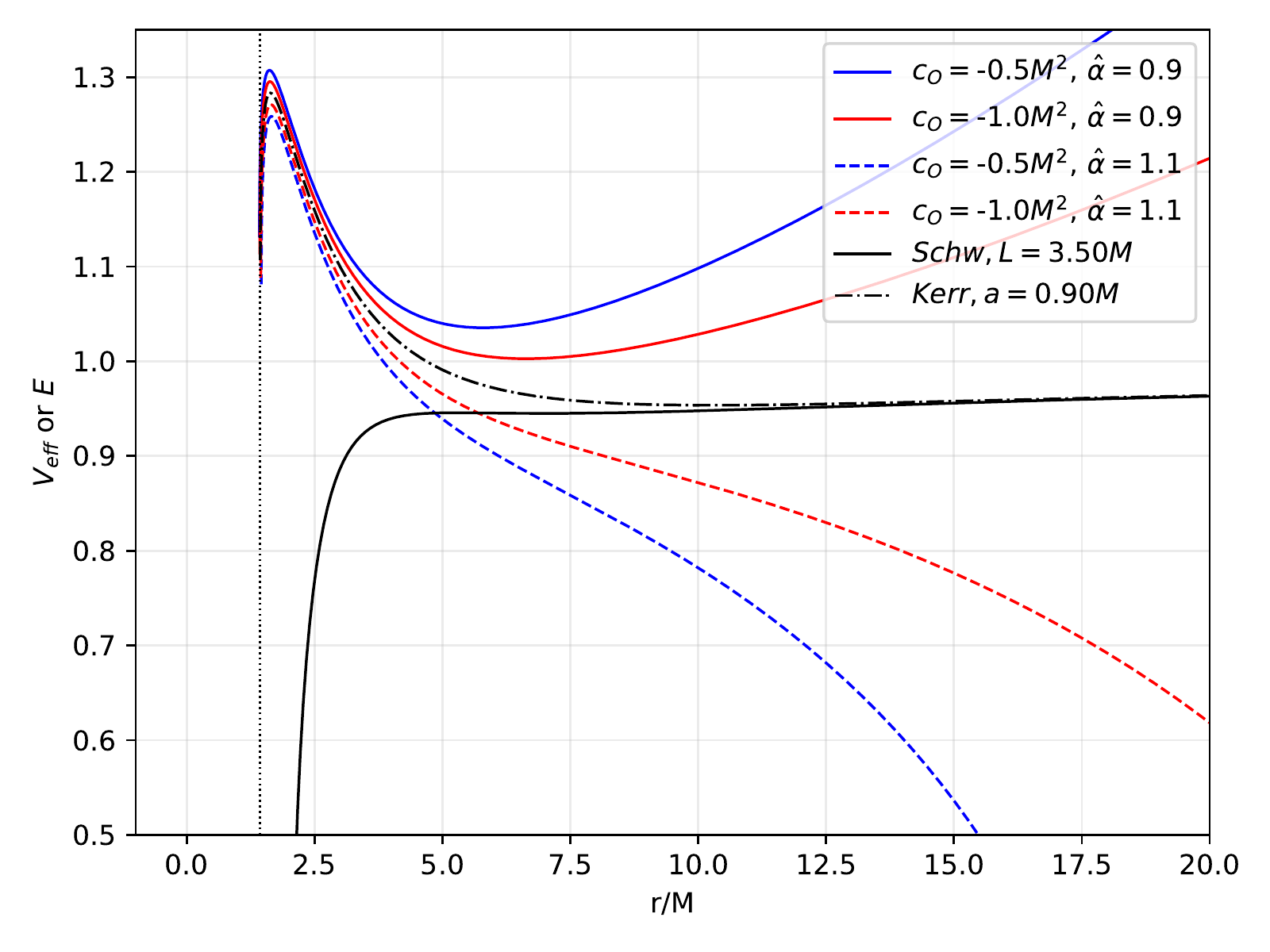} 
    \includegraphics[width=0.48\textwidth]{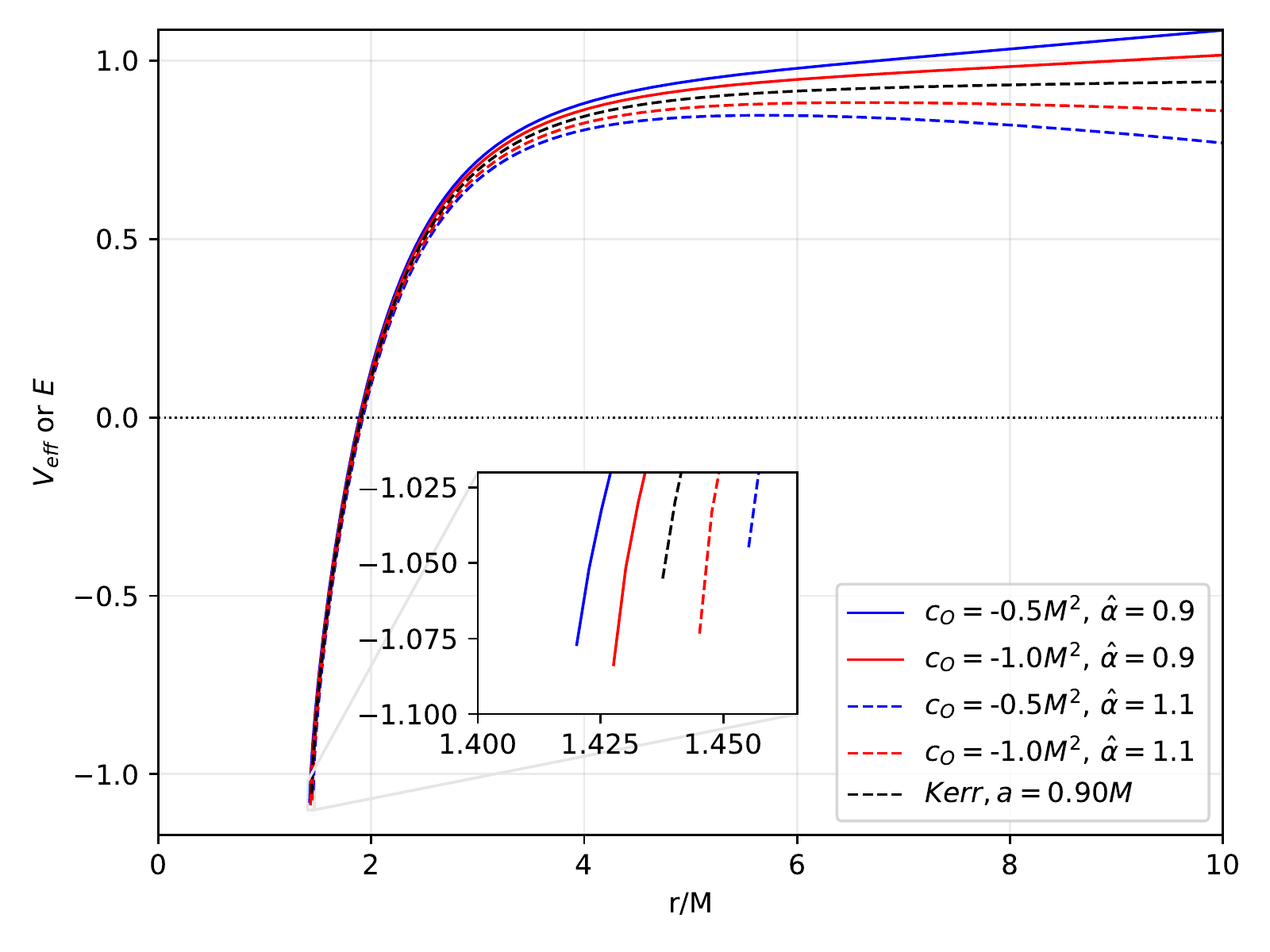}
    \caption{The effective potential as a function of $r/M$ for $L>0$ (left-panel) and $L<0$ (right-panel). In both plots, the black hole spin parameter is set to $a=0.90M$.}
    \label{effpot}
\end{figure*}

To determine the qualitative behavior of  massive particle's motion around the symmergent black hole ({\it i.e.} bound, stable and unstable circular orbits), we use the effective potential formula \cite{Bautista-Olvera2019}
\begin{equation} \label{e44}
    V_{\pm}=\frac{g^{t\phi}}{g^{tt}}L\pm \Bigg\{\left[\left(\frac{g^{t\phi}}{g^{tt}}\right)^2-\frac{g^{\phi \phi}}{g^{tt}}\right]L^2 - \frac{1}{g^{tt}}\Bigg\}^{1/2}.
\end{equation}
written in terms of the inverse metric $g^{\mu\nu}$ and angular momentum $L$. Shown in Fig. \ref{effpot} is the effective potential for $L = \pm 3.50M$.
In this figure, maxima and minima correspond to the unstable and stable circular orbits, respectively. One notes that for  $a=0$ (Schwarzschild) case, the maxima of the effective potential curve are way lower compared to when $a=0.90M$ (Kerr case). We can see that the energy for an unstable circular orbit is higher in the AdS symmergent effect, where less negative $c_{\rm 0}$ giving the highest energy. Its minima, which represents the stable circular orbit, has a greater radius for more negative $c_{\rm 0}$. Peculiar as it may seem, particles with energies higher that than the peak energy tend to get deflected back when it reaches at some radial point near the cosmic horizon (one recalls that we scaled the Symmergent dS and AdS parameters). As for the dS symmergent effect, we can see that the least negative $c_{\rm O}$ gives the lowest peak energy in the unstable circular orbit. Particles with low energy are deflected back to $r\rightarrow \infty$ when they reach low values of $r$. While the AdS type can have elliptical bound orbits the same way as in the Kerr case, we see that dS type does not admit such orbit as far as the value of the angular momentum used herein is concerned. In Fig. \ref{effpot} (right-panel), we plotted also the effective potential in which the particle attains a negative angular momentum. It is known that it does not happen in the Schwarzschild case. As we can see in the right-panel, particles with negative energies can be created near the rotating black hole's event horizon. Clearly, symmergent gravity affects the particle's energy, which in turn contributes to the Penrose process and black hole evaporation.

\begin{figure*}
   \centering
    \includegraphics[width=0.48\textwidth]{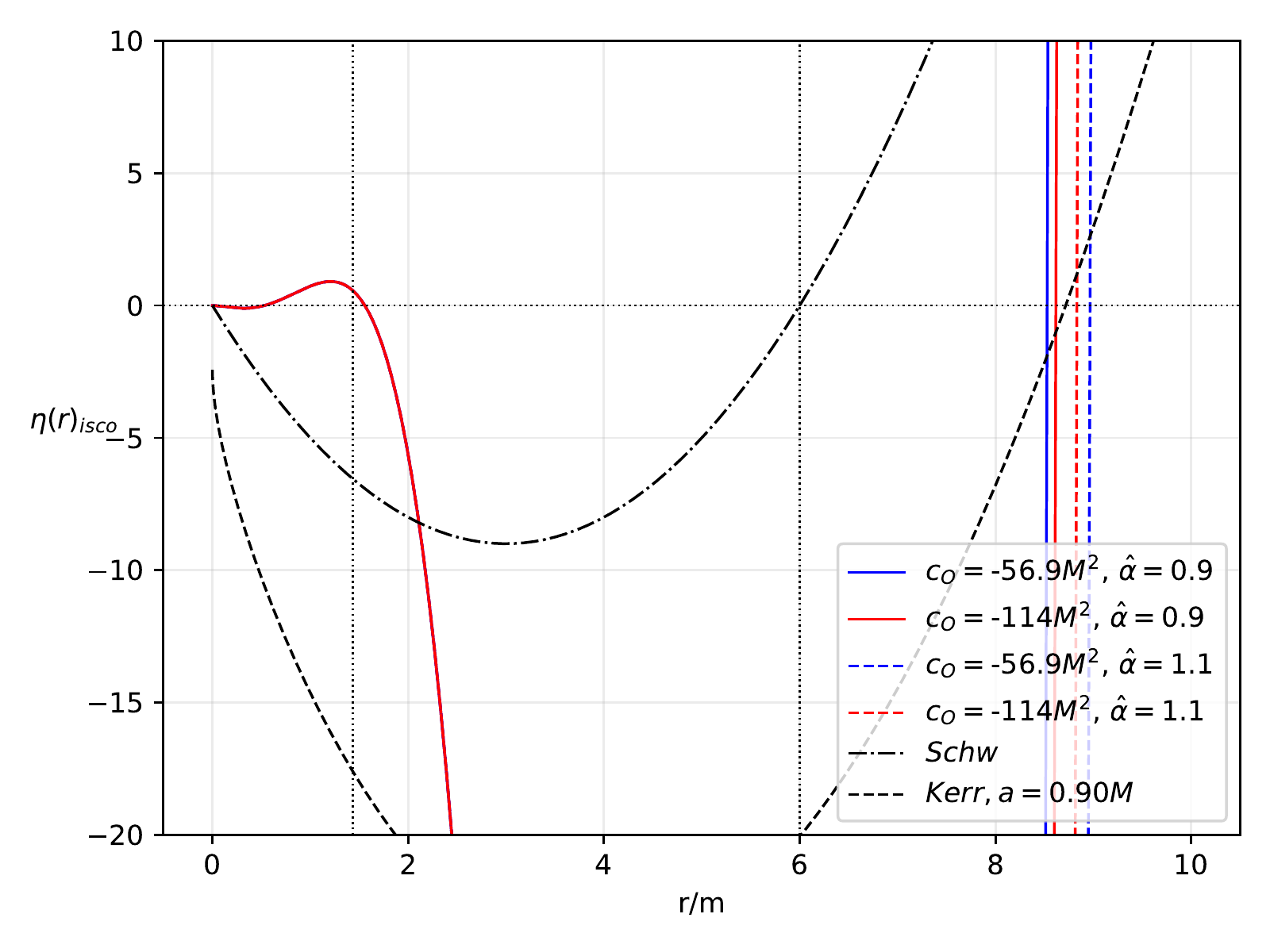} 
    \includegraphics[width=0.48\textwidth]{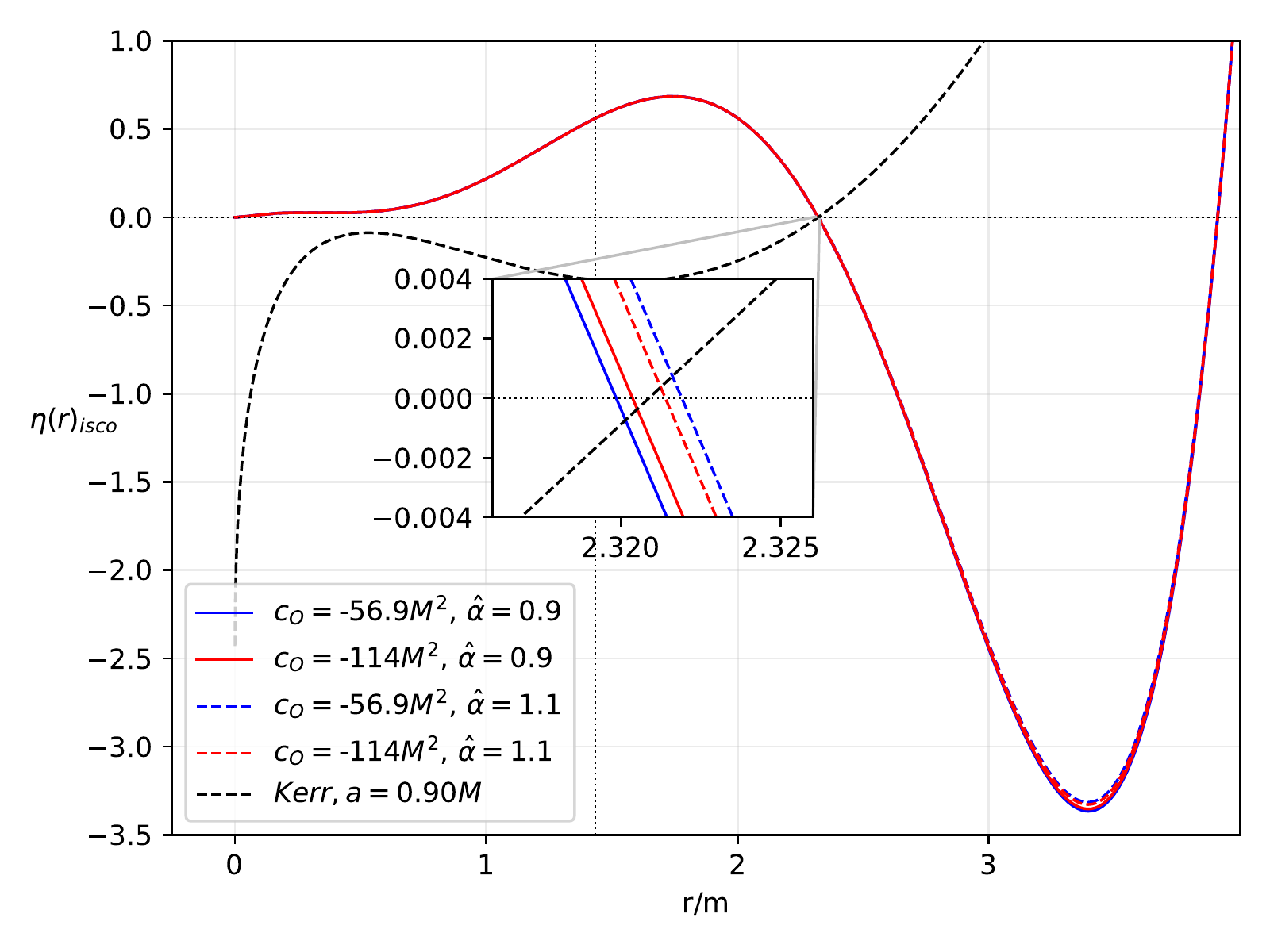}
    \caption{Location of ISCO radius for counter-rotation of a massive particle (left-panel) and co-rotation of a massive particle (right-panel). We set $a=0.90M$ in both panels.}
    \label{isco}
\end{figure*}

The ISCO radius occurs when the maximum and minimum of the effective potential curve merge in an inflection point for a given value of $L$. It is clear that we have two separate cases for ISCO determination since we need to consider both the prograde and retrograde motion of the time-like particles. Besides, such a circular orbit is only marginally stable, which means that any inward perturbation will lead the particle to spiral toward the event horizon. In locating the ISCO radius, we use equation \eqref{e39} and solve for the energy $E_{\text{cir}}$. One starts with \cite{Slany2013,Slany2020},
\begin{equation} \label{e40}
    X^2= L-aE = \frac{r^3 \left(\Delta_r'-2 E_{\text{cir}} ^2 r\right)}{-2 a^2-r \Delta_r'+2 \Delta_r}.
\end{equation}
Differentiating equation \eqref{e40} with respect to $r$ and after some algebra we get ($E_{\text{cir}}=E_{\text{isco}}$)
\begin{equation} \label{e41}
    E_\text{isco}^2=\frac{1}{B r}\biggl\{a^2 \left[-\left(r \Delta_r''+3 \Delta_r'\right)\right]+r \Delta_r \Delta_r''-2 r \Delta_r'^2+3 \Delta_r \Delta_r'\biggr\},
\end{equation}
where $B=-8 a^2+r \left(r \Delta_r''-5 \Delta_r'\right)+8 \Delta_r$. The ISCO radius can then be found by solving the equation \cite{Pantig:2020uhp}
\begin{align} \label{e42}
    \eta(r)_{\text{isco}}&=\pm2 \Delta_r \left(a^2-\Delta_r\right)^2\pm\frac{9}{4} r \Delta_r \left(a^2-\Delta_r\right) \Delta_r'
    \pm\frac{1}{16} r^3 \Delta_r' \left(\Delta_r \Delta_r''-2 \Delta_r'^2\right)\nonumber \\
    &\pm\frac{1}{16} r^2 \bigl[4 \Delta_r \left(a^2-\Delta_r\right) \Delta_r''+\left(15 \Delta_r-4 a^2\right) \Delta_r'^2\bigr] \nonumber \\
    &+a \Delta_r \sqrt{4 a^2+2 r \Delta_r'-4 \Delta_r} \bigl[-a^2+\frac{1}{8} r \bigl(r \Delta_r''-5 \Delta_r'\bigr)+\Delta_r\bigr]=0
\end{align}
which redues to the Kerr case 
\begin{equation} \label{e43}
    3 a^2\mp8 a \sqrt{m} \sqrt{r}+r (6 m-r)=0
\end{equation}
when $\Delta(r)=r^2-2mr+a^2$. The upper and lower signs in equation \eqref{e42} correspond to the prograde and retrograde orbit of the massive particle at the ISCO radius, respectively. We determine ISCO radius numerically  by plotting equation \eqref{e42} in Fig. \ref{isco}.
First, we note that in the Schwarzschild case, the ISCO radius is located at $r=6M$. For the extreme Kerr case, on the other hand, the prograde ISCO is at $r=M$ (which coincides with the prograde photonsphere) and retrograde ISCO is at $r=9M$. Thus, in Fig. \ref{isco}, the left-panel represents the retrograde ISCO where we can see clearly how the symmergent parameters affect the radius. The innermost radius, which determines the accretion disk's innermost region, is smaller in the AdS type whilst the dS type is slightly larger in the Kerr case. We note that the first root of $\eta_\text{isco}$ is unphysical since those radii are smaller than the photonsphere in the retrograde case. Finally, we see the same symmergent effects in the prograde ISCO (right-panel of Fig. \ref{isco}). Before we close this section, we remark that we used lower values of $c_{\rm 0}$ in time-like geodesics than  when  analyzing the null boundaries and geodesics. Thus, there must be some different constraints to be considered in studying time-like geodesics. If we used the same parameters for the null case, the results would be unphysical. Nonetheless, even if $c_{\rm 0}$ are more negative, we see that time-like particles are more sensitive to symmergent effects.

\section{Weak deflection angle from rotating symmergent black holes using Gauss-bonnet theorem} \label{sec5}
In this section, we investigate the deflection angle of light rays from the rotating symmergent black hole in the weak field limit using the methodology given in \cite{Ishihara:2016vdc,Ono:2017pie}. Below, we briefly discuss the finite distance method. First of all, the deflection angle is defined as
 \begin{equation}
 \hat{\theta}=\Psi_\text{R}-\Psi_\text{S}+\phi_\text{RS}
 \label{a1}
 \end{equation}
in which  $\phi_\text{RS}$ is the longitude separation angle, $\Psi_\text{R}$ is the angle between light rays and radial direction, and $\Psi_\text{S}$ is the angle between the observer and the source. 
Unit tangential vector $e^i$  is used to write the above angle as follows
\begin{equation} 
\label{L2}
(e^r,e^\theta,e^\phi)=\epsilon\left(\frac{dr}{d\phi},0,1\right)
\end{equation}
where $\epsilon$ is a radial quantity. For a stationary spacetime  described by the metric 
\begin{equation}
ds^2=-A(r,\theta)dt^2+B(r,\theta)dr^2+C(r,\theta)d\theta^2+D(r,\theta)d\phi^2-2H(r,\theta)dt d\phi,
\end{equation}
one straightforwardly finds 
\begin{equation}
\label{L3}
\epsilon=\frac{A(r)D(r)+H^2(r)}{A(r)(H(r)+A(r)b)},
\end{equation}
where $b$ is the impact parameter. We can write conserved quantities energy  and  the angular momentum per unit mass in the form
\begin{equation}
E= A(r)\dot  t +H(r)\dot \phi \qquad    L= D(r)\dot  \phi-H(r)\dot t,
\label{labe}
\end{equation}
where the overdot denotes the derivative of the coordinates $t$ and $\phi$ with respect to the affine parameter $\lambda$ (proper time). In the equatorial plane ($\theta=\frac{\pi}{2}$), the line  element reduces to
\begin{equation}
dl^2\equiv \gamma_{ij}dx^{i}dx^{j}
\label{labe1}
\end{equation}
with the spatial metric $\gamma_{ij}$. It is then possible to define 
\begin{equation}
\label{L1a}
\cos\Psi\equiv \gamma_{ij}e^iR^j.
\end{equation}
as the angle between the direction $e^i$ and the radial vector $R^j=(\frac{1}{\sqrt{\gamma_{rr}}},0,0)$. Having $\cos\Psi$ at hand, we  use equations (\ref{L2}) and (\ref{L1a}) to get 
\begin{equation}
\sin\Psi=\frac{H(r)+A(r)b}{\sqrt{A(r)D(r)+H^2(r)}}.
\label{EE}
\end{equation}

Now, using this methodology we can calculate the deflection angle of the symmergent black hole in the weak field limit. Indeed, with the impact parameter $b=\frac{L}{E}$ and radial coordinate $1/u$ in place of $r$ we get 
\begin{eqnarray}
\left( \frac{du}{d\phi}\right)^{2} &=&\frac{1}{b^2}-\frac{4 a M u}{b^3}-u^2+2 M u^3+\frac{1-\hat{\alpha}}{24\pi c_{\rm 0}}-\frac{2 a }{3 b^3 u^2}\frac{(1-\hat{\alpha})}{8\pi c_{\rm 0}}-\frac{8 a  M}{3 b^3 u}\frac{(1-\hat{\alpha})}{8\pi c_{\rm 0}}+\mathcal{O}\left[M^2,\left(\frac{1-\hat{\alpha}}{8\pi c_{\rm 0}}\right)^2,a^2\right]
\end{eqnarray}
where we considered only the retrograde solution. For the prograde solution, we just  flip the sign of the spin parameter $a$. It is convenient to start by computing the separation angle integral. This can be done by writing $ \left( \frac{du}{d\phi}\right)^{2} =F(u)$ and extracting the angle
\begin{equation}
\phi_\text{RS} = \int^\text{R}_\text{S} d\phi= \int^{u_0}_{u_\text{S}}\frac{1}{\sqrt{F(u)}}du +\int^{u_0}_{u_\text{R}}\frac{1}{\sqrt{F(u)}}du , 
\end{equation}
where $u_\text{S}$ ($u_\text{R}$) is inverse distance to the source (observer), and $u_0$ is the inverse of the closest approach $r_{0}$. Considering the  weak field and slow rotation approximations, the impact parameter can be related to $u_{0}$ as
\begin{equation}
b = \frac{1}{u_{0}}+M-2aM u_{0}+\mathcal{O}\left[M^2,\left(\frac{1-\hat{\alpha}}{8\pi c_\text{0}}\right)^2,a^2\right].
\end{equation}

Then, performing the necessary calculations we find
\begin{eqnarray}
\phi_\text{RS} &=& \phi_\text{RS}^\text{Kerr}
 + \left(\frac{ u_\text{R}}{ \sqrt{1-b^2 u_\text{R}^2}}+\frac{ u_\text{S}}{ \sqrt{1-b^2 u_\text{S}^2}}\right)\frac{b^3}{6}\frac{(1-\hat{\alpha})}{8\pi c_\text{0}} +\left(\frac{b \left(2-3 b^2 u_\text{R}^2\right)}{2 \left(1-b^2 u_\text{R}^2\right){}^{3/2}}+\frac{b \left(2-3 b^2 u_\text{S}^2\right)}{2 \left(1-b^2 u_\text{S}^2\right){}^{3/2}}\right)\frac{M}{3}\frac{(1-\hat{\alpha})}{8\pi c_\text{0}} \notag \\ &+& \left( \frac{1-2 b^2 u_\text{R}^2}{u_\text{R} \sqrt{1-b^2 u_\text{R}^2}}+\frac{1-2 b^2 u_\text{S}^2}{u_\text{S} \sqrt{1-b^2 u_\text{S}^2}}\right) \frac{a}{3}\frac{(1-\hat{\alpha})}{8\pi c_\text{0}}+ \mathcal{O}\left[M^2,\left(\frac{1-\hat{\alpha}}{8\pi c_\text{0}}\right)^2,a^2\right],
\end{eqnarray}
where the Kerr term is given by
\begin{eqnarray}
 \phi_\text{RS}^\text{Kerr} &=& \phi_\text{RS}^\text{SW} - \left(\frac{1}{\sqrt{1-b^2 u_\text{R}^2}}+\frac{1}{ \sqrt{1-b^2 u_\text{S}^2}}\right)\frac{2aM}{b^2}.
\end{eqnarray}
In this equation,   the Schwarzschild   term  is \begin{eqnarray}\phi_\text{RS}^\text{Schw}=\pi -\arcsin\left(b u_\text{R}\right)-\arcsin\left(b u_\text{S}\right)+ \left(\frac{2-b^2 u_\text{R}^2}{ \sqrt{1-b^2 u_\text{R}^2}}+\frac{2-b^2 u_\text{S}^2}{\sqrt{1-b^2 u_\text{S}^2}}\right) \frac{M}{b}.\end{eqnarray}
To get the remaining terms appearing in the light deflection angle expression,   one should identify the $\Psi$ terms. We find 
\begin{eqnarray}
\sin \Psi&=&b u-b M u^2+2 a M u^2- \left(\frac{b M}{6} - \frac{a  }{3 u}+\frac{b }{6 u}\right)\frac{(1-\hat{\alpha})}{8\pi c_\text{0}} +\mathcal{O}\left[M^2,\left(\frac{1-\hat{\alpha}}{8\pi c_\text{0}}\right)^2,a^2\right].
\end{eqnarray}
This relation  produces 
\begin{eqnarray}
\Psi_\text{R}&-&\Psi_\text{S}=\Psi_\text{R}^\text{Kerr}-\left( \frac{1}{u_\text{R} \sqrt{1-b^2 u_\text{R}^2}}+\frac{1}{u_\text{S} \sqrt{1-b^2 u_\text{S}^2}}\right) \frac{b }{6}\frac{(1-\hat{\alpha})}{8\pi c_\text{0}}-\left( \frac{2 b^2 u_\text{R}^2-1}{\left(1-b^2 u_\text{R}^2\right){}^{3/2}}+\frac{2 b^2 u_\text{S}^2-1}{\left(1-b^2 u_\text{S}^2\right){}^{3/2}}\right) \frac{b M }{6}\frac{(1-\hat{\alpha})}{8\pi c_\text{0}}\notag \\ &+&\left( \frac{1}{u_\text{R} \sqrt{1-b^2 u_\text{R}^2}}+\frac{1}{u_\text{S} \sqrt{1-b^2 u_\text{S}^2}}\right) \frac{a }{3}\frac{(1-\hat{\alpha})}{8\pi c_\text{0}}+  \mathcal{O}\left[M^2,\left(\frac{1-\hat{\alpha}}{8\pi c_\text{0}}\right)^2,a^2\right],
\end{eqnarray}
where one has 
\begin{eqnarray}
\Psi_\text{R}^\text{Kerr}&-&\Psi_\text{S}^\text{Kerr}=\Psi_\text{R}^\text{Schw}-\Psi_\text{R}^\text{Schw}+\left( \frac{u_\text{R}^2}{\sqrt{1-b^2 u_\text{R}^2}}+\frac{u_\text{S}^2}{\sqrt{1-b^2 u_\text{S}^2}}\right) 2 a M,
\end{eqnarray}
and where  one has found 
\begin{eqnarray}
\Psi_\text{R}^\text{Schw}-\Psi_\text{R}^\text{Schw}=\left( \arcsin\left(b u_\text{R}\right)+\arcsin\left(b u_\text{S}\right)-\pi\right) -\left( \frac{u_\text{R}^2}{\sqrt{1-b^2 u_\text{R}^2}}+\frac{u_\text{S}^2}{\sqrt{1-b^2 u_\text{S}^2}}\right)M b .
\end{eqnarray}
Combining the above equations, we get  an expression of  the  light deflection angle given by 

\begin{eqnarray} \label{enear}
\hat{\theta}&=& \left(\sqrt{1-b^2 u_\text{R}^2}+\sqrt{1-b^2 u_\text{S}^2} \right)\frac{2 M}{b}- \left( \sqrt{1-b^2 u_\text{R}^2}-\sqrt{1-b^2 u_\text{S}^2} \right)\frac{2 a M}{b^2}\notag \\ &-& \left( \frac{1-b^2 u_\text{R}^2}{u_\text{R} \sqrt{1-b^2 u_\text{R}^2}}+\frac{1-b^2 u_\text{S}^2}{u_\text{S} \sqrt{1-b^2 u_\text{S}^2}}\right) \frac{b}{6}\frac{(1-\hat{\alpha})}{8\pi c_\text{0}}+\left( \frac{1}{\sqrt{1-b^2 u_\text{R}^2}}+\frac{1}{\sqrt{1-b^2 u_\text{S}^2}}\right)\frac{b  M}{6}\frac{(1-\hat{\alpha})}{8\pi c_\text{0}} \notag \\ &+& \left(\frac{\sqrt{1-b^2 u_\text{R}^2}}{u_\text{R}}+\frac{\sqrt{1-b^2 u_\text{S}^2}}{u_\text{S}}\right) \frac{2 a }{3}\frac{(1-\hat{\alpha})}{8\pi c_\text{0}}+\mathcal{O}\left[M^2,\left(\frac{1-\hat{\alpha}}{8\pi c_\text{0}}\right)^2,a^2\right] .
\end{eqnarray}
This form can be reduced to a simplified one using certain convenable approximations. Taking $u_\text{S}b<<1$ and $u_\text{R}b<<1$,     we can get an expression involving divergent terms coupled to  the cosmological   contributions. These terms should be existed to show the cosmological background dependence. The desired deflection angle of light rays  is found to be 
\begin{eqnarray} \label{efar}
\hat{\theta} &=& \frac{4 M}{b}-\frac{4 a M}{b^2} +\frac{b  M}{3}\frac{(1-\hat{\alpha})}{8\pi c_\text{0}}-\left(\frac{1}{u_\text{R}}+\frac{1}{u_\text{S}}\right)\frac{b }{6}\frac{(1-\hat{\alpha})}{8\pi c_\text{0}} +\left(\frac{1}{u_\text{R}}+\frac{1}{u_\text{S}}\right)\frac{2 a }{3}\frac{(1-\hat{\alpha})}{8\pi c_\text{0}}+\mathcal{O}\left[M^2,\left(\frac{1-\hat{\alpha}}{8\pi c_\text{0}}\right)^2,a^2\right].
\end{eqnarray}
An examination shows that this expression recovers many previous findings.   Without the cosmological contributions, we get the results of the ordinary Symmergent black hole solutions investigated in \cite{Symmergent-bh2}.
\begin{figure}
   \centering
    \includegraphics[width=0.48\textwidth]{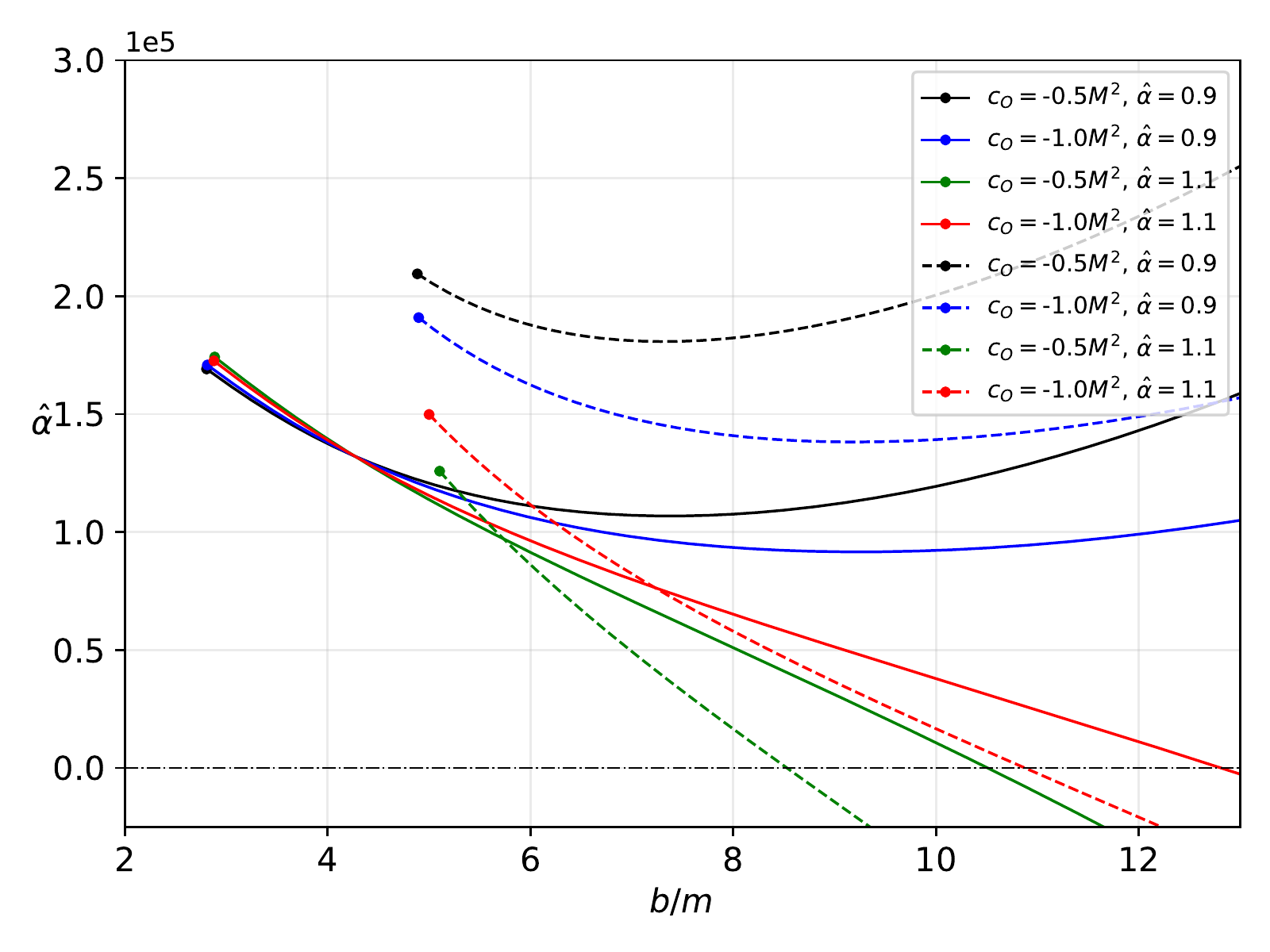}
    \includegraphics[width=0.48\textwidth]{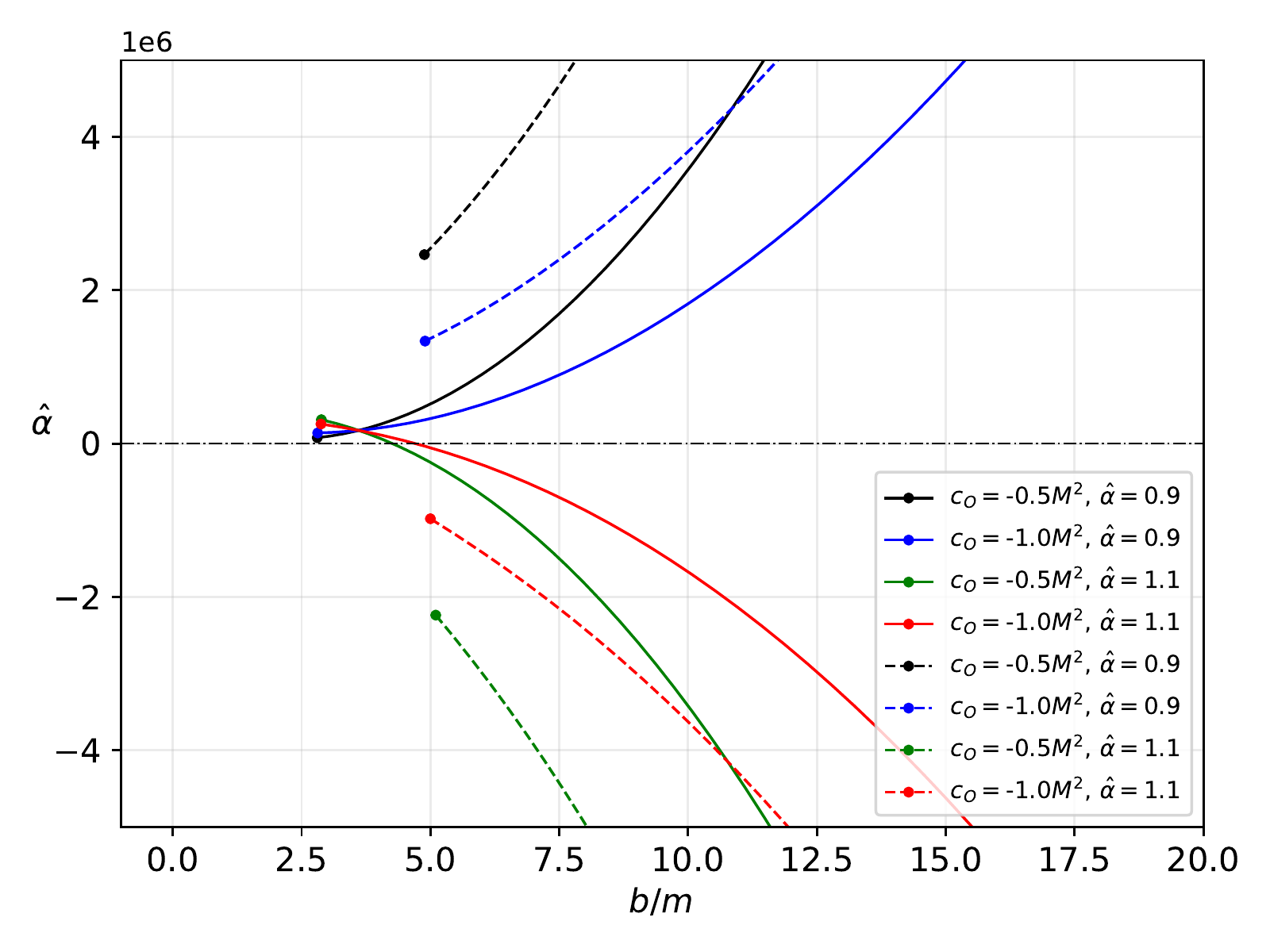} 
    \caption{Weak deflection angle of photons as a function of $b/m$ for various values of the spin and symmergent parameters. The solid lines correspond to $a>0$ (prograde) and dashed lines correspond to $a<0$ (retrograde). In making these plots we take $u_\text{S} = u_\text{R} = 0.5b^{-1}$ (left), $u_\text{S} = u_\text{R} = 0.01b^{-1}$ (right), and $\hat{\alpha}=0.90$. The horizontal dotted lines represent critical impact parameters in each case.}
    \label{wda}
\end{figure}

Plotted in Fig. \ref{wda} are the equations \eqref{enear} (left-panel) and \eqref{efar} (right-panel). The solid curves represent photons co-rotating with the black hole, while the dashed curves represent the counter-rotating photons. For the left-panel, which holds for finite distance, the results show that the co-rotating photons produce weak deflection angle at lower impact parameters compared to the counter-rotating photons. If we compare $\hat{\theta}$ for a specific orbital direction of photons, we see that the symmergent AdS type gives a higher value for $\hat{\theta}$ at large impact parameters. However, AdS type in the counter-rotation case gives a higher $\hat{\theta}$ than the co-rotation case. Finally, the right-panel stands for the case where we used the large distance approximation. Since the symmergent parameters are scaled, the location of the receiver in this case is comparable to the cosmic horizon. In this case, the symmergent effect is strong, giving the effects shown in the plot. We could see that there are cases where $\hat{\theta}$ is repulsive. In general, weak deflection angle is seen to have a strong sensitivity to the symmergent  parameters.

\subsection{Particle acceleration near rotating Symmergent black hole backgrounds}\label{sec6}

As Banados, Silk and West (BSW) show that collisions in equatorial plane near a Kerr black hole can occur with an arbitrarily high center of mass-energy and Kerr black holes can serve as particle
accelerators \cite{Banados:2009pr}. Then the BSW mechanism has been studied different black hole spacetimes \cite{Halilsoy:2015rna,Halilsoy:2015qta,Yang:2012we,Li:2010ej,Zhang:2018ocv,Wang:2022qmg,Dymnikova:2022sre}.
In the present work, we find that the symmergent parameter $c_{\rm 0}$ has an important effect on the result. To do that first we consider the particle motion on the equatorial plane ($\theta = \frac{\pi}{2}$, $\rho^2 = r^2$). In fact, generalized momenta $P_\mu$ can be written as ($\mu, \nu = t, r, \phi, \theta$)
\begin{equation}
P_\mu=g_{\mu \nu}\dot{x}^\nu
\end{equation}
where the dot stands for  derivative with respect to the affine parameter $\lambda$. Then, one can write the generalized momenta $P_t$ (test particle's energy per unit mass $E$) and $P_\phi$ (the angular momentum per unit mass $L$ parallel to the symmetry axis)
\begin{eqnarray}
P_t &=& g_{tt} \dot{t} + g_{t \phi} \dot{\phi},\label{4-veloE} \\
P_\phi &=& g_{\phi \phi} \dot{\phi} + g_{t \phi} \dot{t}\label{4-veloL}
\end{eqnarray}
where one keeps in mind that $P_t$ and $P_\phi$ are constants of motion. 

Our goal is to study the CM energy of the two-particle collision in the background spacetime of the rotating symmergent black holes. To this end, we start with  the CM energy $E_{\text{CM}}$  \cite{Banados:2009pr}
\begin{eqnarray}
E_{\text{CM}}^{2}=2 m_{0}^{2}\left(1-g_{\mu \nu} u_{1}^{\mu} u_{2}^{\nu}\right) \end{eqnarray}
in which $u^{\mu}_1, u^{\nu}_2$ are the 4-velocity vectors of the two particles ($u=(\dot{t},\, \dot{r},\, 0,\, \dot{\phi})$). Correspondingly, particle $i=1,2$ has angular momentum per unit mass $L_i$ and energy per unit mass $E_i$. We also consider  particles with the same rest mass $m_0$. Initially the two particles are at rest at infinity ($E_1/m_0 = 1$ and $E_2/m_0 = 1$) and then they approach the black hole and collide at a distance $r$. Then, their CM energy  takes the form
 \begin{equation}
E_{\text{CM}}^2=\frac{2 m_0^2}{\Delta_r r^{2}}\left[\left(r^{2}+a^{2}\right)^{2}-a\left(L_{1}+L_{2}\right)\left(r^{2}+a^{2}-\Delta_r\right)+L_{1} L_{2}\left(a^{2}-\Delta_r\right)+\Delta_r\left(r^{2}-a^{2}\right)-X_{1} X_{2}\right] \label{Ecm}
\end{equation}
with the functions 
 \begin{equation}
X_{i}=\sqrt{\left(a L_{i}-r^{2}-a^{2}\right)^{2}-\Delta_r\left(\left(L_{i}-a\right)^{2}+\mu^{2} r^{2}\right)}.
\end{equation}

\begin{figure}
   \centering
    \includegraphics[width=0.48\textwidth]{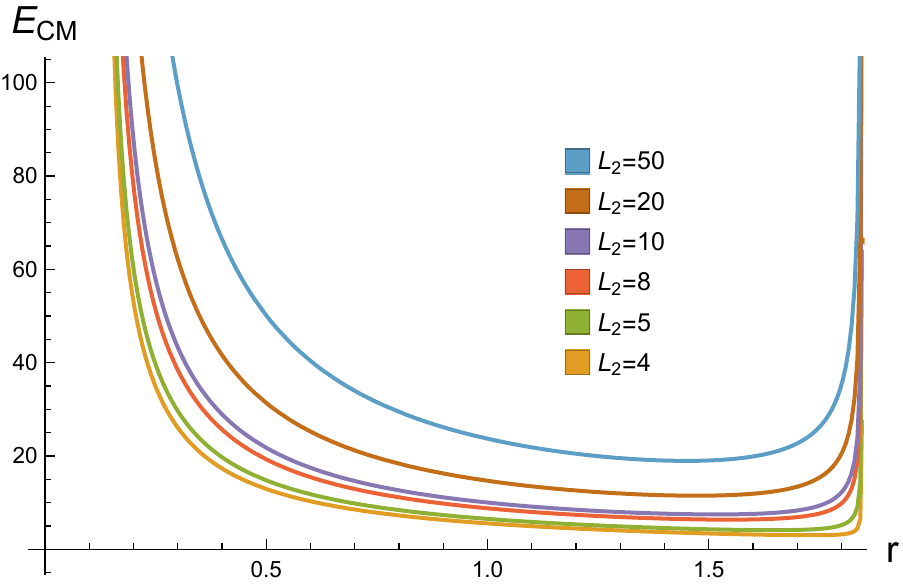}
    \includegraphics[width=0.48\textwidth]{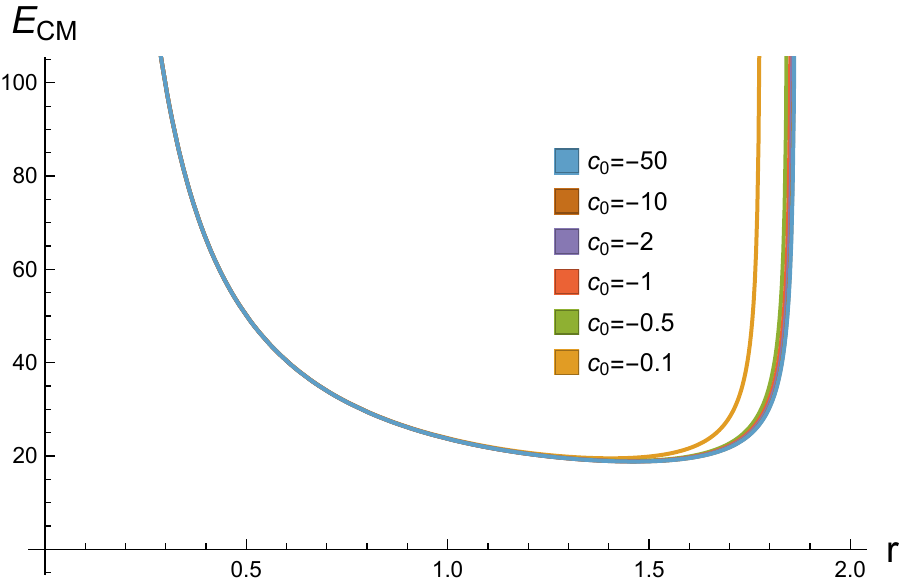}
    \caption{Dependence of the CM energy $E_{\text{CM}}$ on the radial distance $r$ for different values of the spin and symmergent parameters. We consider two parameter sets $m=M=1$, $L_1=-4$,$c_0=-0.5$, $\hat{\alpha}=0.9$ and $a=0.5$ (left-panel) as well as $m=M=1$, $L_1=-4$,$L_2=50$, $\hat{\alpha}=0.9$ and $a=0.5$ (right-panel). It is clear that the CM energy blows up at the horizons.}  \label{BSW}
\end{figure}

Collisions occur at the horizon of the black hole so that  $\Delta_r=0$ in equation (\ref{Ecm}) so CM energy could diverge when the particles approach the horizon. It is not difficult to see that the denominator of $E_{\text{CM}}^2$ is zero there. Evidently,
the maximal energy of the collision occurs if $L_1$ and $L_2$ are opposite (such as head-on collision). Plotted in Fig. \ref{BSW} is $E_{\text{CM}}$ as a function of the radial distance $r$. As is seen from the figure, CM energy blows up at the horizons. It demonstrates the radial dependence of the CM energy  of the particles moving along circular orbits with the different symmergent parameters. The symmergent parameter decreases the CM energy.

\section{Conclusion} \label{sec7}

In this paper, we have studied rotating black holes in symmergent gravity, and used deviations from the Kerr black hole to constrain the parameters of the symmergent gravity. Symmergent gravity generates parameters of the gravitational sector from flat spacetime loops. It induces the gravitational constant $G$ and quadratic-curvature coefficient $c_{\rm O}$.
In the limit in which all fields are degenerate in mass, the vacuum energy $V_{\rm O}$ can be expressed in terms of $G$ and $c_{\rm O}$. We parametrize deviation from this degenerate  limit by the parameter ${\hat \alpha}$. The black hole  spacetime is dS for ${\hat \alpha} < 1$ and AdS for  ${\hat \alpha} > 1$. In constraining the symmergent parameters $c_{\rm O}$ and ${\hat \alpha}$, we utilize the EHT observations on the M87* and Sgr. A* black holes.

We have analyzed symmergent gravity parameters for different spinning black hole properties.  We first investigated the effects of symmergent gravity on the photonsphere, black hole shadow, and certain observables related to them. Our findings show that  location of the photonsphere depends on the symmergent effects due to its coupling to the black hole spin parameter. We have seen how the instability of the photon radii depends on and gets affected by $c_{\rm O}$ and $\hat{\alpha}$. Also, symmergent gravity effect is evident, even for observers deviating from the equatorial plane. Interesting effects are also seen in the energy emission rate, which may directly affect the black hole's lifetime. Constraints on symmergent parameters from the shadow radius data coming from recent observations are also discussed. For a low and high spin parameter $a$ values (the parameter $\hat{\alpha}$ dictates whether the symmergent effect mimics the dS and AdS type), we found that the parameter  $c_{\rm 0}$ fits within $\pm2\sigma$ at $95\%$ confidence level for M87* than in Sgr. A*. It means that the true value is somewhere within such confidence level. The plot in Fig. \ref{sha_cons} reveals that the detection of symmergent effects can be achieved more easily in M87* than in Sgr. A*, for the obvious reason that the farther the galaxy is the stronger the symmergent effects are. Finally, for comparison, we also briefly analyzed such an effect on the time-like geodesic. Our findings indicate that massive particles are more sensitive to deviations caused by the symmergent parameter than null particles.

In addition, we investigated the symmergent effects on photons' weak deflection angle as they traverse near the black hole. We also considered the finite distance effects. We have found  that the deflection angle depends on $\hat{\alpha}$ but the deviations are also caused by how far the receiver is from the black hole. This effect is clearly shown in Fig. \ref{wda}. It  implies that the weak deflection angle can detect symmergent effects more easily for black holes that are significantly remote compared to our location.

Lastly, we study the particle collision near the rotating symmergent black hole background and we analyze the possibility that rotating symmergent black holes could act as particle accelerators. We show that rotating symmergent black holes can serve as particle accelerators because the center-of-mass energies blow up at the horizons. As a result, the BSW mechanism depends on the value of the Symmergent parameter $c_\text{0}$ of the black hole.

Future research direction may include investigation of the shadow, and showing its dependence on the observer's state. One may also study the spherical photon orbits or the stability of time-like orbits.

\acknowledgments 
We thank to conscientious referee for their comments, questions and criticisms. We thank Xiao-Mei Kuang for useful discussions. A. \"O. acknowledges hospitality at Sabanc{\i} University where this work was completed. A. {\"O}. and R. P. would like to acknowledge networking support by the COST Action CA18108 - Quantum gravity phenomenology in the multi-messenger approach (QG-MM).

\bibliography{apj}
\end{document}